\documentclass[letterpaper,preprintnumbers,prd,onecolumn,nofootinbib,nobibnotes,showpacs]{revtex4}
\usepackage{amsfonts}
\usepackage{mathrsfs}
\usepackage{epsfig}
\usepackage{graphicx}%
\usepackage{dcolumn}
\usepackage{amsmath}
\newcommand{\beq}{\begin{equation}}
\newcommand{\eeq}{\end{equation}}
\newcommand{\bea}{\begin{eqnarray}}
\newcommand{\eea}{\end{eqnarray}}
\newcommand{\bseq}{\begin{subequations}}
\newcommand{\eseq}{\end{subequations}}

\makeatletter
\def\btt#1{\texttt{\@backslashchar#1}}%
\DeclareRobustCommand\bblash{\btt{\@backslashchar}}%
\makeatother
\usepackage{color}
\usepackage{amssymb}

\begin{document}

\title{Non-minimal kinetic coupled gravity: inflation on the Warped DGP brane}
\author{F. Darabi$^{1,2}$\thanks{Email: f.darabi@azaruniv.edu}, A. Parsiya$^1$ \thanks{Email: a.parsiya@azaruniv.edu} and K. Atazadeh$^1$\thanks{Email: k.atazadeh@azaruniv.edu}\\
{\small $^1$Department of Physics, Azarbaijan Shahid Madani University, Tabriz 53714-161, Iran}\\
{\small $^2$Research Institute for Astronomy and Astrophysics of Maragha (RIAAM), Maragha 55134-441, Iran}}
\date{\today}
\begin{abstract}
 We consider the non-minimally kinetic coupled version of DGP brane model, where the kinetic term of the scalar field is coupled to the metric and Einstein tensor on the brane by a coupling constant $\zeta$. We obtain the corresponding field equations, using the Friedmann-Robertson-Walker metric and the perfect fluid, and study the inflationary scenario to confront the numerical analysis of six typical  scalar field potentials with the current observational results. We  find that among the suggested potentials and coupling constants,    subject to the e-folding $N=60$,
the potentials $V(\phi)=\sigma\phi$,  $V(\phi)=\sigma\phi^2$ and $V(\phi)=\sigma\phi^3$
provide the best fits with both Planck+WP+highL data and Planck+WP+highL+BICEP2 data.
\end{abstract}

\maketitle

\section{Introduction}
During the last several years, the braneworld scenario has been considerably
studied within the variety of different models. According to this scenario, we are living over a three-dimensional hypersurface in a higher-dimensional spacetime; the standard model particles are confined on the brane, and the gravitons propagate in the bulk spacetime
\cite{early,Arkani-Hamed}. Using this scenario in a five-dimensional spacetime, Randall and
Sundrum (RS) proposed two kinds of resolutions for the hierarchy problem \cite{Randall_Sundrum}. They showed that
in a five-dimensional spacetime one may derive the effective Einstein equations for the $4D$ brane metric obtained by projecting the $5D$ metric onto the braneworld which result in the most general form of the $4D$ gravitational field equations for a braneworld observer \cite{SMS,SSM,MW,M_supple}. Induced gravity brane model proposed by Dvali, Gabadadze and Porrati (DGP) is another example of this scenario to account for the self accelerating behaviour of the
universe \cite{Dvali}.
Many authors have studied the geometrical \cite{b-i-gravity,b-i-gravity-pert,b-i-gravity-disc,b-i-gravity-bh,b-i-gravity-6dim}   as well as the cosmological \cite{B-i-cosmology,B-i-obs-cosmology,B-i-cosmology2}
aspects of this new $5D$ gravitational model.
DGP model with a bulk cosmological
constant and a tension of the brane, with energy scale much larger than the $5D$ Planck mass,  leads to the effective cosmological constant on the brane which is extremely reduced in contrast to  the RS model, even if the cosmological constant and the tension are not fine-tuned
\cite{EGE}.

The inflationary scenario  can
resolve the problems of  standard cosmology such as the flatness, horizon, monopole and relics problems. In most of the successful inflationary models the  universe is filled with a  scalar field so called
{\it Inflaton} whose potential energy is dominant over its kinetic energy\cite{1,2,3,4,5,6,7,8}.
However,  several
problems remain without  concrete solutions \cite{6,7,8,9}. Hence, many other inflationary
models such as the braneworld models \cite{10,11,12,13,14}, models with
non-minimally coupled inflaton field \cite{15,16,17,18,19,20,21,22}, modified gravity
\cite{23,24,25}, and models with a wide range of potentials  have attracted so much attention in the recent
years. In this regard, variety of models have been proposed
however those models are viable that show consistency with observational
data and provide us with a mechanism for generating the initial fluctuations
and perturbations in the early universe as the seeds for the
formation of the structures in the universe. In such models,
the fluctuations in the scalar field as well as the transverse and traceless parts of the metric lead to the scalar and tensor power spectrum,
respectively \cite{1,2,3,4,5,6,7,8}. The scalar power spectrum
is nearly scale-invariant, with the order of unity, and the good point is
that the exact value of spectral index can be
obtained by using the observational data. Moreover, the running of spectral index and  the tensor-to-scalar ratio can also be constrained observationally.
Comparison between the calculated values of these parameters and the recent  observational data are the most powerful probes for ruling out or keeping a specific inflation model. 

Such a study has already been done in the context of non-minimal DGP braneworld inflation in Ref.\cite{KNNR}, where the non-minimal feature of the model was attributed to the non-minimal coupling between the {\it inflaton field} and the  induced {\it Ricci scalar} on the brane. Also, the observational constraint were analyzed with respect to the background of Planck+WMAP9+BAO data and the potential $V(\phi)\sim \exp(-\beta\phi)$ was
obtained as the best fit case.  

Here, we develop a study similar to Ref.\cite{KNNR} to find other possible inflation
models, consistent with the observations, in DGP braneworld scenario. However, the present study is much different from Ref.\cite{KNNR} in four senses. The first is that here we follow a different approach to use the non-minimal feature in our model. Rather than considering the non-minimal coupling between the {\it inflaton field} and the  induced {\it Ricci scalar} on the brane, we consider a non-minimal coupling between the {\it kinetic term of the inflaton field} and {\it Einstein tensor} on the brane. Such models are known as {\it non-minimal kinetic coupled gravity} \cite{KC}. The second is that here we analyze our observational constraint with respect to the background data of Planck+WP+highL+BICEP2 rather than Planck+WMAP9+BAO. The third is that here we consider six types of inflaton potentials, more or less different from those of Ref.\cite{KNNR}, and perform a numerical analysis on the inflationary parameters of this model to confront them with Planck+WP+highL+BICEP2 data. The forth
is that here we obtain three new best fit potentials rather than (and different
from) one
obtained in Ref.\cite{KNNR}. It was already
found that some potentials which are suitable for inflation in 4-dimensional model, cannot lead to a successful inflation in the minimal case of 5-dimensional
DGP model. Moreover, some potentials which are not compatible with observational data in a 4-dimensional model, can lead to viable results in a minimal 5-dimensional DGP model \cite{KNNR}.
In this paper, by considering a non-minimal coupling between the kinetic term of the scalar field and
Einstein tensor in a 5-dimensional DGP model, we obtain new scalar field potentials  suitable for inflation. 
\\

\section{ Field equations in the brane scenario }

We assume a $5D$ bulk spacetime $({\cal M},{}^{(5)}g_{AB})$ with the coordinates $X^A ~(A=0,1,2,3,5)$ and a $4D$ brane $(M,g_{\mu\nu})$  located at a hypersurface ${\cal B} (X^A) = 0$.
The standard action for the braneworld is written as 
\begin{eqnarray}
\label{action}
S =  S_{\rm Bulk}+ S_{\rm brane},
\end{eqnarray}

\begin{eqnarray}
S_{\rm Bulk} =\int_{\cal M} d^5X \sqrt{-{}^{(5)}g} \left[
{1 \over 2 \kappa_5^2} {}^{(5)}R +
{}^{(5)}L_{\rm m} \right],
\label{bulk_action}
\end{eqnarray}

\begin{eqnarray}
S_{\rm brane}=\int_{M} d^4 x\sqrt{-g}
\left[  {1\over\kappa_5^2} K^\pm
+ L_{\rm brane}(g_{\alpha\beta},\psi)
\right],
\label{brane_action}
\end{eqnarray}
where $\kappa_5^2$ corresponds to the $5D$ gravitational constant $m_5$, ${}^{(5)}L_{\rm m}$
and ${}^{(5)}R$ are the $5D$ the matter
Lagrangian in the bulk and scalar curvature, respectively.
Also, $x^\nu ~(\nu=0,1,2,3)$ are the induced $4D$  coordinates on
the  brane, $K^\pm$ is the trace of extrinsic
curvature on either side of the brane \cite{GH,ChaRea99} and
$L_{\rm brane}(g_{\alpha\beta},\psi)$ is the effective $4D$ Lagrangian which is given by a typical functional
of the brane metric $g_{\alpha\beta}$ and matter fields
$\psi$.

The five-dimensional Einstein equations in the bulk are given by
\begin{eqnarray}
{}^{(5)}G_{AB} = \kappa_5^2 \,\, \left[ \,{}^{(5)}T_{AB}
+\tau_{AB}\, \delta({\cal B})\,\right]
\, ,
\label{5dEinstein}
\end{eqnarray}
where \begin{eqnarray}
{}^{(5)}T_{AB} &\equiv &-2 {\delta {}^{(5)}\!L_{\rm m} \over \delta
{}^{(5)}g^{AB}}  +{}^{(5)}g_{AB}{}^{(5)}\!L_{\rm m}~,
\label{em_tensor_of_bulk}\,
\end{eqnarray}
is the energy-momentum tensor of bulk matter fields, and 
\begin{eqnarray}
\tau_{\mu\nu}\equiv -2 {\delta L_{\rm
brane} \over \delta
g^{\mu\nu}}  +g_{\mu\nu}L_{\rm
brane}     \,,
\label{em_tensor_of_brane}
\end{eqnarray}
is the effective energy-momentum tensor localized by $\delta({\cal B})$ on the brane.
The induced $4D$ metric can be written as $g_{AB} = {}^{(5)}g_{AB} - n_{A}n_{B},$
where $n_A$ is the spacelike unit-vector field normal to the brane hypersurface $M$.
Following \cite{SMS},   \cite{M_supple}, and \cite{b-i-gravity-pert} one obtains the gravitational field equations on the braneworld as \cite{effective}

\begin{eqnarray}
G_{\mu\nu}
&=& {2 \kappa_5^2 \over 3}\biggl[{}^{(5)}T_{RS}~
g^{R}_{~\mu}  g^{S}_{~\nu}
+ g_{\mu\nu} \biggl({}^{(5)}T_{RS}~n^R n^S
-{1 \over
4}{}^{(5)}T\biggr)
     \biggr]+\kappa_5^4\pi_{\mu\nu} - E_{\mu\nu}\, ,
\label{eq:effective}
\end{eqnarray}
%
\begin{eqnarray}
D_\nu \tau^{~\nu}_{\mu}=-2  \,{}^{(5)}T_{RS} \, n^R g^{S}_{~\mu}\,,
\label{Codazzi2}
\end{eqnarray}
where
\begin{equation}
\pi_{\mu\nu}=
-\frac{1}{4} \tau_{\mu\alpha}\tau_\nu^{~\alpha}
+\frac{1}{12}\tau\tau_{\mu\nu}
+\frac{1}{8}g_{\mu\nu}\tau_{\alpha\beta}\tau^{\alpha\beta}
-\frac{1}{24}
g_{\mu\nu}\tau^2,
\label{pidef}
\end{equation}
and
\begin{equation}
E_{\mu\nu} = {}^{(5)}C_{MRNS}~n^M n^N
g_{~\mu}^{R}~ g_{~\nu}^{S} .
\label{Edef}
\end{equation}

\section{DGP brane's  Model with non-minimal kinetic coupled gravity}
\label{III}

In this section, we modify DGP braneworld model by a non-minimal kinetic coupling  term in
the  Lagrangian
\begin{eqnarray}\label{brane_action2}
L_{\rm brane}=  {\mu^2 \over 2} R-\frac{1}{2}( g^{\mu\nu}-\zeta G^{\mu\nu})\nabla_\mu \phi \nabla_\nu \phi -V(\phi)-  \lambda + L_{\rm m}  \, ,
\end{eqnarray}
where $\mu$ is a mass scale which may correspond to the four dimensional Planck mass $m_4$,
$R$ is the Ricci scalar, $\zeta$ is a coupling parameter with dimension of $(length)^2$, $V(\phi)$ is the scalar  field potential, $\lambda$ is the
tension of the brane, and $L_m$ is the Lagrangian of other matters on
the brane.
 The presence of Einstein tensor in the kinetic term of the inflaton field
is novel and casts this model in the context of non-minimal coupled gravity. Also, we take only
a cosmological constant ${}^{(5)}\Lambda$ in the bulk.

\subsection{ Field Equations on the brane}
To obtain the field equations on the brane, we calculate the
energy-momentum tensor of the brane  
as\begin{eqnarray}
\tau^{\mu}_{~\nu}=-\lambda \delta^{\mu}_{~\nu} +T^{\mu}_{~\nu}-\mu^2
G^{\mu}_{~\nu}+\Omega^{\mu}_{\nu}+\zeta \Theta^{\mu}_{\nu},
\end{eqnarray}
where
\begin{equation}
\Omega^{\mu}_{\nu}={\nabla^{\mu}\phi\nabla_\nu\phi
-\frac{{1}}{2}\delta^\mu_{\nu}\nabla_\rho\phi\nabla^\rho\phi}-\delta^\mu_{\nu}V(\phi),
\end{equation}
\begin{eqnarray}\label{energy}
 \Theta_{\mu\nu}&=&-{\textstyle\frac12}\nabla_\mu\phi\,\nabla_\nu\phi\,R
+2\nabla_{\alpha}\phi\,\nabla_{(\mu}\phi R^{\alpha}_{\nu)}
+\nabla^\alpha\phi\,\nabla^\beta\phi\,R_{\mu\alpha\nu\beta}
+\nabla_\mu\nabla^\alpha\phi\,\nabla_\nu\nabla_\alpha\phi
-\nabla_\mu\nabla_\nu\phi\,\square\phi\;
\\ \nonumber
&&-{\textstyle\frac12}(\nabla\phi)^2
G_{\mu\nu}
+g_{\mu\nu}\big[-{\textstyle\frac12}\nabla^\alpha\nabla^\beta\phi\,\nabla_\alpha\nabla_\beta\phi
+{\textstyle\frac12}(\square\phi)^2
-\nabla_\alpha\phi\,\nabla_\beta\phi\,R^{\alpha\beta}
\big].
\nonumber\\
\end{eqnarray}
Substituting this equation into
Eq.(\ref{eq:effective}),
one can find the effective equations for $4D$ metric
$g_{\mu\nu}$ as \cite{EGE}
\begin{eqnarray}
\left(1+{\lambda\over 6}\kappa_5^4\mu^2\right)
G_{\mu\nu}+\kappa_5^4\mu^2 {\cal
K}_{\mu\nu\rho\sigma}(T_{\alpha\beta})G^{\rho\sigma}+\Lambda g_{\mu\nu}
={\lambda\over 6}\kappa_5^4 T_{\mu\nu} +\kappa_5^4\left[
\pi_{\mu\nu}^{(T)} +\mu^4 \pi^{(G)}_{\mu\nu}\right]-E_{\mu\nu},
\end{eqnarray}
where
\begin{equation}
{\cal K}_{\mu\nu\rho\sigma}=
{1\over 4}\left(g_{\mu\nu}T_{\rho\sigma}-g_{\mu\rho}T_{\nu\sigma}
-g_{\nu\sigma}  T_{\mu\rho}\right)
+{1\over 12}\Bigl[T_{\mu\nu}g_{\rho\sigma}
+T\left(g_{\mu\rho}g_{\nu\sigma}
-g_{\mu\nu}g_{\rho\sigma}\right)\Bigr],
\end{equation}
\begin{equation}
\Lambda={1\over 2}\left[{}^{(5)}\Lambda+{1\over 6}\kappa_5^4
\lambda^2\right],
\end{equation}
\begin{equation}
\pi_{\mu\nu}^{(T)}=
-\frac{1}{4} T_{\mu\alpha}T_\nu^{~\alpha}
+\frac{1}{12}TT_{\mu\nu}
+\frac{1}{8}g_{\mu\nu}T_{\alpha\beta}T^{\alpha\beta} -\frac{1}{24}
g_{\mu\nu}T^2,
\end{equation}
\begin{equation}
\pi_{\mu\nu}^{(G)}=
-\frac{1}{4} G_{\mu\alpha}G_\nu^{~\alpha}
+\frac{1}{12}GG_{\mu\nu}
+\frac{1}{8}g_{\mu\nu}G_{\alpha\beta}G^{\alpha\beta}-\frac{1}{24}
g_{\mu\nu}G^2,
\end{equation}
 $T$ and $G$ being the  trace of energy-momentum and Einstein  tensors, respectively.
Because of the Bianchi identity, the Codazzi equation reads as $D^\nu \tau_{\mu\nu}=0$ which implies
the energy momentum conservation, i.e.
\begin{eqnarray}
D^\nu T_{\mu\nu}=0\, .
\end{eqnarray}

\subsection{Cosmology of non-minimal kinetic coupled DGP model}
 We take the spatially flat isotropic and homogeneous FRW line element on the brane
\begin{equation}
\label{8} ds^{2}=-dt^{2}+a^{2}(t)\delta_{ij}dx^{i}dx^{j}\,,
\end{equation}
where $\delta_{ij}$ is a symmetric 3-dimensional metric and $a(t)$ is the scale factor.  Studying such universe with a perfect fluid and following \cite{SMS}, we can write
$D^\nu \pi_{\mu\nu}=0$
implying that
\begin{eqnarray}
D^\nu E_{\mu\nu}=0.
\label{eq:Emn}
\end{eqnarray}
The original field equations (\ref{eq:effective}) can be
written as \cite{EGE}
\begin{eqnarray}
G^0_{~0} &=& -{1\over 2} {}^{(5)}\!\Lambda +\kappa_5^4 \pi^0_{~0}- E^0_{~0},
\label{eq:Friedmann00}
\\
G^i_{~j} &=& -{1\over 2} {}^{(5)}\!\Lambda\delta^i_{~j} +\kappa_5^4
\pi^i_{~j}- E^i_{~j},
\label{eq:Friedmann0}
\end{eqnarray}
where
\begin{eqnarray}
G^0_{~0} &=& -3\left(H^2 +{k\over a^2}\right), \nonumber \\
G^i_{~j} &=& -\left(2\dot{H}+3 H^2 +{k\over a^2}\right)\delta^i_{~j},
\end{eqnarray}
and
     \begin{eqnarray}
\pi^0_{~0} &=& -{1\over 12} \left(\tau^0_{~0}\right)^2,
\nonumber \\
\pi^i_{~j} &=& {1\over 12}
\tau^0_{~0}\left(\tau^0_{~0}-2\tau^1_{~1}\right)
\delta^i_{~j},
\end{eqnarray}
with
\begin{eqnarray}
\tau^0_{~0} = -(\lambda+\rho) -\mu^2 G^0_{~0}
+\frac{1}{2}(1+9\zeta H^2)\dot{\phi}^2-V(\phi),
\end{eqnarray}
\begin{eqnarray}
\tau^i_{~j} = (P-\lambda) \delta^i_{~j}-\mu^2 G^i_{~j}
+\dot{\phi}^2 \left[1+\zeta\left(2\dot{H}+3H^2+\frac{\kappa}{a^2}
+4H\ddot{\phi}\dot{\phi}^{-1}\right)\right]\delta^i_{~j}
-V(\phi)\delta^i_{~j}.
\end{eqnarray}
Eqs.(\ref{eq:Friedmann00}) and (\ref{eq:Friedmann0}) are written as \cite{EGE}
\begin{eqnarray}
3X={1\over 2} {}^{(5)}\Lambda +E^0_{~0}
+{\kappa_5^4\over 12}
\left(-\lambda -\rho  +3\mu^2X
+\frac{1}{2}(1+9\zeta H^2)\dot{\phi}^2-V(\phi)
\right)^2,
\label{H_const}
\end{eqnarray}
\begin{eqnarray}
&&\left[1+{\kappa_5^4\over 6}\left(-\lambda -\rho +3\mu^2X
+\frac{1}{2}(1+9 \zeta H^2)\dot{\phi}^2-V(\phi\right)
\left(\mu^2+\frac{\zeta}{2}\dot{\phi}^2\right)\right]Y
\nonumber\\
&=&-{2\over 3}E^0_{~0}
+{\kappa_5^4\over 12}\left(\rho+P+9\zeta H^2\dot{\phi}^2
-2\zeta H \dot{\phi}\ddot{\phi}+\frac{3}{2}\dot{\phi}^2X\right)
\left(-\lambda -\rho +3\mu^2X
+\frac{1}{2}(1+9\zeta H^2)\dot{\phi}^2-V(\phi)\right),
\label{dynamical_eq}
\end{eqnarray}
where
    \begin{eqnarray}
X&=&H^2+{k\over a^2},
\nonumber \\
Y&=&\dot{H}-{k\over a^2}.
\end{eqnarray}
By using Eq.(\ref{eq:Emn}), we can obtain the  equation of motion for $E^0_{~0}$ as
follows
\begin{eqnarray}
\dot{E}^0_{~0}+4HE^0_{~0}=0.
\end{eqnarray}
By integrating from the above equation we can easily find  
\begin{eqnarray}\label{e0}
E^0_{~0}={{\cal E}_0\over a^4},
\end{eqnarray}
where ${\cal E}_0$ is an integration constant.
Now, we must solve the equation (\ref{H_const}),
 as a quadratic equation with respect to
$X$, which can be written as
    \begin{eqnarray}
H^2+{k\over a^2}={1\over 3\mu^2}\Bigl[\,\rho_m+\rho_0\bigl(1
+ \epsilon{\cal A}(\rho, a)\bigr)\,\Bigr]
\label{Friedmann_Dvali1}
\, ,
\end{eqnarray}
where
 \begin{eqnarray}
\rho_m=\rho+\rho _{\phi}=\rho+\frac{1}{2}(1+9\zeta H^2)\dot{\phi}^2+V(\phi),
\end{eqnarray}
and
\begin{eqnarray}
\rho_0=m_\lambda^4+6{\kappa_5^{-4}\over \mu^2}\,,
\end{eqnarray}
with the mass scale $m_\lambda=\lambda^{1/4}$.
Also, $\epsilon$ stands for  either $+1$ or $-1,$ and ${\cal A}$ is defined by \cite{EGE}
\begin{eqnarray}
{\cal A}&\equiv& \left[{\cal A}_0^2+{2\Gamma\over
\rho_0}\left(\rho_m-\mu^2
{{\cal E}_0\over a^4}
\right)\right]^{1\over 2},
\end{eqnarray}
where
\begin{eqnarray}
{\cal A}_0&=&\sqrt{1-2\Gamma{\mu^2\Lambda\over \rho_0}},\\
\Gamma&=6&{m_5^{6}\over \rho_0\mu^2}
~~~(0<\Gamma\leq 1)\,.
\end{eqnarray}
Equation (\ref{Friedmann_Dvali1}) is considered as the Friedmann equation of this model.
Note that the choice of sign for $\epsilon$ has a geometrical meaning \cite{B-i-cosmology} and it is determined by the initial condition of the universe.

{\section{Inflation}}
In the slow-roll regime, we have $9 \zeta H^2\dot{\phi}^{2}\ll V(\phi), \ddot{\phi}\ll3H\dot{\phi}$ \cite{Reh} and because of \eqref{e0} at inflationary stage ($a\gg{\cal E}_0$), we may ignore the integration constant ${\cal E}_0$ by setting ${\cal E}_0=0$. So,
the energy density takes the following form
\begin{equation}
\label{11} \rho_m\approx V(\phi)\equiv V\,.
\end{equation}
 By varying the Lagrangian (\ref{brane_action2}) with respect to the scalar field we  have
\beq\label{fieldeq3}
  (\ddot\phi+3H\dot\phi)+3\zeta(H^2\ddot\phi
  +2H\dot{H}\dot\phi+3H^3\dot\phi)+V'=0.
\eeq
 Using the slow-roll approximation,  we obtain the following equation of motion
\begin{equation}
\label{10}
{3H\dot{\phi}}\left(1+\zeta \big(2\dot{H}+3H^2\big)\right)+V'=0.
\end{equation}
So, the Einstein equations in slow-roll approximation take the following forms, respectively
as\begin{eqnarray}
H^2={1\over 3\mu^2}\Bigl[\,V(\phi)+\rho_0\bigl(1
+ \epsilon{\cal A}\bigr)\,\Bigr]
\label{Friedmann_Dvali2}
\, ,
\end{eqnarray}
\begin{eqnarray}
\dot{H}=-{1\over 2\mu^2}\frac{{V'}^2}{9H^2\big(1+\zeta(2\dot{H}+3H^2)\big)}\Bigl
[\,1+\epsilon\Gamma {\cal A}^{^{-1}}\,\Bigr]
\label{Friedmann_Dvali3}
\, ,
\end{eqnarray}
where
\begin{eqnarray}
{\cal A}&\equiv& \left[{\cal A}_0^2+2\Gamma{V\over
\rho_0}\right]^{1\over 2},
\end{eqnarray}
and ${'}$ denotes $\frac{d}{d\phi}$.
The slow-roll parameters defined by $\varepsilon\equiv
-\frac{\dot{H}}{H^{2}}$ and $\eta\equiv
\varepsilon-\frac{\dot{\varepsilon}}{2H\varepsilon}$ take the
following forms, respectively as
\begin{eqnarray}
\varepsilon={\mu^2\over2 }\frac{V'^2}{V^2} \frac{1}{[1+\zeta(2\dot{H}+3H^2)]}\frac{\Bigl
[\,1+\epsilon\Gamma {\cal A}^{^{-1}}\,\Bigr]}{\Bigl[\,1+\frac{\rho_0}{V}\bigl(1
+ \epsilon{\cal A}\bigr)\,\Bigr]^2}
\label{varepsilon}
\, ,
\end{eqnarray}

and
\begin{equation}
\eta = \frac{\varepsilon-\frac{1}{6H^2\big(1+\zeta(2\dot{H}+3H^2)\big)}\bigg(-2\big(V''-\frac{V'^2}{V}\big)-V'\big(\frac{{
\cal B}^{'}}{\cal B}-2\frac{{
\cal C}^{'}}{\cal C}\big)\bigg)-\bigg(\frac{{3\zeta
\varepsilon H^2}}{1+\zeta(2\dot{H}+3H^2)}\bigg)}{1-\bigg(\frac{2\zeta
\varepsilon H^2}{1+\zeta(2\dot{H}+3H^2)}\bigg)},
\end{equation}
where \begin{eqnarray} \label{A}
{\cal A}&=& \left({\cal A}_0^2+2\Gamma{V\over
\rho_0}\right)^{1\over 2},
\nonumber\\
{\cal B}&=& 1+\epsilon\Gamma{\cal A}^{-1},
\nonumber\\
{\cal C}&=&1+\frac{\rho_0}{V}\bigl(1
+ \epsilon{\cal A}\bigr)\,.
\end{eqnarray}
 The number of e-folding  is given by
$N=\int_{t_{i}}^{t_{f}} H dt$ where ($t_{i}$) and ($t_{e}$) are the initial and end time of inflation, respectively. For a warped DGP model with a
non-minimally kinetic coupled gravity on the brane,  we will get the following expression
\begin{eqnarray}\label{N}
N=-\frac{1}{\mu^2}\int_{\phi_{i}}^{\phi_{f}}
\Bigg(\frac{V}{V'}\Bigg){\bigg[1+\zeta(2\dot{H}+3H^2)\bigg]}{\bigg(1+\frac{\rho_0}{V}\bigl(1
+ \epsilon{\cal A}\bigr)\,\bigg)}d\phi ,
\end{eqnarray}
where $\phi_{i}$ and $\phi_{f}$ are the values of
$\phi$ when the radius of universe  crosses the Hubble horizon during
inflation and exits the inflationary phase, respectively.
A useful tool to test the viability of inflationary models is the spectrum of perturbations produced due to the quantum fluctuations
around their homogeneous background values.  The
conformal-Newtonian metric is given by \cite{27,28,29}
\begin{equation}
ds^{2}=a^{2}(\eta)\bigg[-\big(1+2\Psi\big)d\eta^{2}+\big(1-2\Psi\big)\delta_{ij}\,dx^{i}dx^{j}\bigg],\label{13}
\end{equation}
 where $\Psi$ is called the \textit{Bardeen potential}.
  The primordial power spectrum is defined by the following expression \cite{Basset}
\begin{eqnarray}
 P_{_{R}}=\left(\frac{H^2}{2\pi \dot{\phi}}\right)^2\bigg|_{k=aH}.
\end{eqnarray}
 Using the scalar field equation of motion in slow-roll regime (i.e.
equation(\ref{10})), we  get
\begin{eqnarray}\label{pr1}
 P_{_{R}}=\left(\frac{H^2}{2\pi \dot{\phi}}\right)^2\bigg|_{k=aH}=\frac{V'}{12\pi^2\mu^6}
\Bigg(\frac{V}{V'}\Bigg)^3{\bigg[1+\zeta(2\dot{H}+3H^2)\bigg]^2}{\Bigr[1+\frac{\rho_0}{V}\bigl(1
+ \epsilon{\cal A}\bigr)\,\Bigr]^3}.
\end{eqnarray}
In the slow-roll regime, we know that $\varepsilon\ll1$ and $\dot{H}\simeq 0$
; so the equation (\ref{pr1}) takes the following approximate form
\begin{eqnarray}\label{pr3}
 P_{_{R}}\simeq\frac{V'}{12\pi^2\mu^6}
\Bigg(\frac{V}{V'}\Bigg)^3{\Bigr(1+3\zeta H^2\Bigr)^2}{\Big(1+\frac{\rho_0}{V}\bigl(1
+ \epsilon{\cal A}\bigr)\,\Big)^3}.
\end{eqnarray}The scalar
spectral index, which describes the scale-dependence of the
perturbations is defined as
\begin{equation}
n_{s}-1=\frac{d \ln P_{_{R}}^{2}}{d \ln k}\,,\label{14}
\end{equation}
where $d \ln k(\phi)=d N(\phi)$.
As is seen, for $n_{s}=1$ the power spectrum of the perturbation is scale invariant. In our warped
DGP model, we obtain the scalar spectral index in the slow-roll
regime as follows
\begin{eqnarray}\label{15}
n_{s}-1=-\frac{\mu^2}{2}
\Big(\frac{V'}{V}\Big)^2\bigg({\frac{3(1+\epsilon\Gamma{\cal A}^{-1})}{1+\frac{\rho_0}{V}(1
+ \epsilon{\cal A})}-\frac{2V''V}{V'^2}}\bigg)\bigg(\big(1+\zeta(2\dot{H}+3H^2)\big)\big(1+\frac{\rho_0}{V}\bigl(1
+ \epsilon{\cal A}\bigr)\big)\bigg)^{-1} ,
\end{eqnarray}
where ${\cal A}$ is defined in (\ref{A}).
The running of spectral index in our model is given by
\begin{eqnarray}\label{alpha}
{\alpha}&=&\frac{d \ln n_{_{s}}}{d \ln k}
\nonumber\\
&=&\frac{1}{\big[1+\zeta(2\dot{H}+3H^2\big]{\cal C}}\Bigg\{-\mu^2\bigg(\frac{V'}{V}\bigg)\frac{1}{1+\zeta(2\dot{H}+3H^2)}
\bigg(\frac{{\cal C}^{'}}{\cal {C}}+\frac{\zeta {\cal C} {V}}{\mu^2\big(1+\zeta(2\dot{H}+3H^2)\big)}\big(\frac{V'}{V}+\frac{{\cal
C}^{'}}{\cal C}\big)\bigg)
\nonumber\\
&-&6\bigg[{\mu^2}
\big(\frac{V'}{V}\big)^2\times\big({\mu^2}
\frac{V^{''}}{V}\big)
-4\bigg(\frac{\mu^2}{2}
\Big(\frac{V'}{V}\Big)^2\bigg)^2\bigg]\frac{{\cal B}}{{\cal
C}}
+\frac{\big(\mu^2\frac{V'}{V}\big)}{\big(1+\zeta(2\dot{H}+3H^2)\big){\cal C}}\times6 \bigg(\frac{\mu^2}{2}
\Big(\frac{V'}{V}\Big)^2\bigg)\times\Big(\frac{{\cal B}'{\cal C}-{\cal
C}'{\cal B}}{{\cal C}^2}\Big)
\nonumber\\
&+&2\bigg(-{\mu^2}
\Big(\frac{V'}{V}\Big)^2\bigg({\mu^2}
\frac{V^{''}}{V}\bigg)
+\mu^4\frac{V' V''}{V^2}\bigg)\Bigg\},
\end{eqnarray}
where
\begin{eqnarray}
{\cal A}&=& \left({\cal A}_0^2+2\Gamma{V\over
\rho_0}\right)^{1\over 2},
\nonumber\\
{\cal B}&=& 1+\epsilon\Gamma{\cal A}^{-1},
\nonumber\\
{\cal C}&=&1+\frac{\rho_0}{V}\bigl(1
+ \epsilon{\cal A}\bigr)\,.
\end{eqnarray}
The tensor perturbation (gravitational wave) amplitude of a given mode, at the time of
Hubble crossing, is given by
\begin{equation}
P_{_{T}}^{2}={\frac{8}{\mu^2}}\bigg(\frac{H}{2\pi}\bigg)^2\Bigg|_{k=aH}\,.\label{17}
\end{equation}
In our model and in the slow-roll regime, we find the primordial tensor perturbation
 \begin{equation}
P_{_{T}}^{2}=\frac{8}{(2\pi)^2}\frac{V}{3\mu^2}\bigg(1+\frac{\rho_0}{V}\bigl(1
+ \epsilon{\cal A}\bigr)\bigg),\label{17-1}
\end{equation}
 and  the tensor  spectral index is given by
\begin{equation}
n_{_{T}}=\frac{d \ln P_{_{T}}^{2}}{d \ln k}\,,\label{19}
\end{equation}
so, in the slow-roll regime we can
express it as follows
\begin{equation}
n_{_{T}}=-2\varepsilon\,.\label{20}
\end{equation}
 Now, we evaluate the tensor-to-scalar ratio as
\begin{equation}\label{21}
r\equiv\frac{P_{_{T}}^{2}}{P_{_{R}}^{2}}\simeq\frac{\mu^2}{2}
\Big(\frac{V'}{V}\Big)^2\frac{16}{\big(1+\zeta(2\dot{H}+3H^2)\big)^2[\,1+\frac{\rho_0}{V}\bigl(1
+ \epsilon{\cal A}\bigr)\,]^2} .
\end{equation}

Up to now, we have presented the equations of cosmological dynamics. In the
following,
 we perform numerical analysis on the inflationary parameters of the warped DGP
model with a non-minimally kinetic coupled gravity  on the brane. Now, we shall consider
some types of potentials by substituting them in the integral of equation
(\ref{N}) and then solve this equation. But, first we should find the value of $\phi$
at the end of inflation, namely $\phi_{f}$, by setting $\varepsilon=1$ in Eq.(\ref{varepsilon})  (corresponding to the end of inflation). Then,
we put it in (\ref{N}) and  find
$\phi_{i}$ in terms of $N$  and  substitute $\phi_{i}$ in $n_{s}$, $r$ and $\alpha$, for any given values of $N$. Now, we can compare these
inflationary parameters with the recent observational data.\\

\section{Observational constraint}
In this section, first we introduce our model parameters  as  
\begin{equation}
{}~~\kappa_5^2\sim m_5^{-3},~~\mu\sim m_{4}~~,~~\lambda\sim0~~,~~{}^{(5)}\Lambda\sim
0~~,~~~~\Gamma\sim0.99~~
,~~k\sim0.002Mpc^{-1}~~,
\end{equation}
together with two parameters ($\zeta, \sigma$);
the first one is the nonminimal coupling constant for which we shall take different
values so that one can decide which one shows best fit with observations
and the second one is the parameter defined in the scalar field potential (se bellow) which,
with no loss of generality, we will take its value to be of the order of unity. Then, we  investigate these models and compare the results with the
observational
data. 
\subsection{$V(\phi)={\sigma}\phi^{\frac{1}{2}}$}
 
\begin{figure}
\includegraphics[width=8.55 cm]{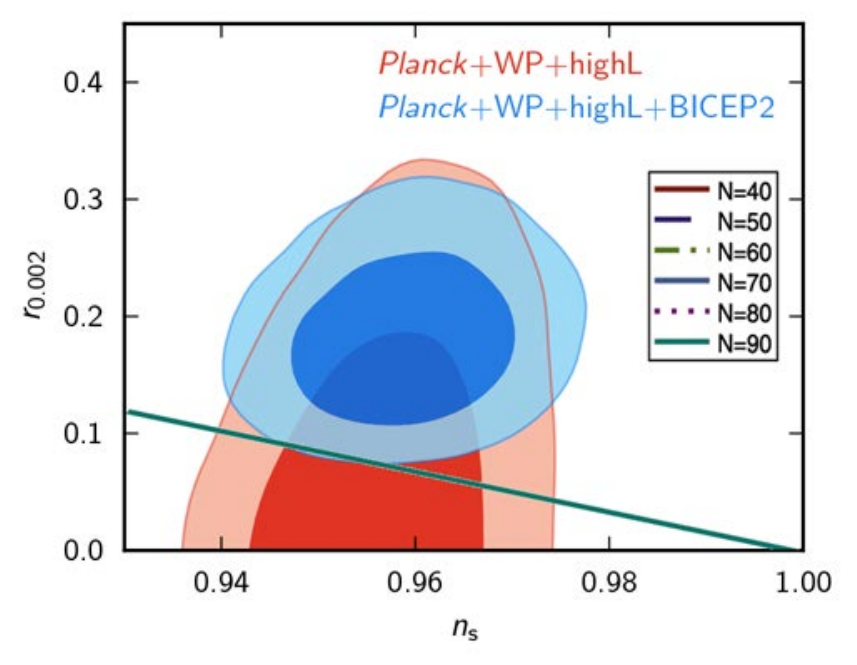}
\includegraphics[width=8.8 cm]{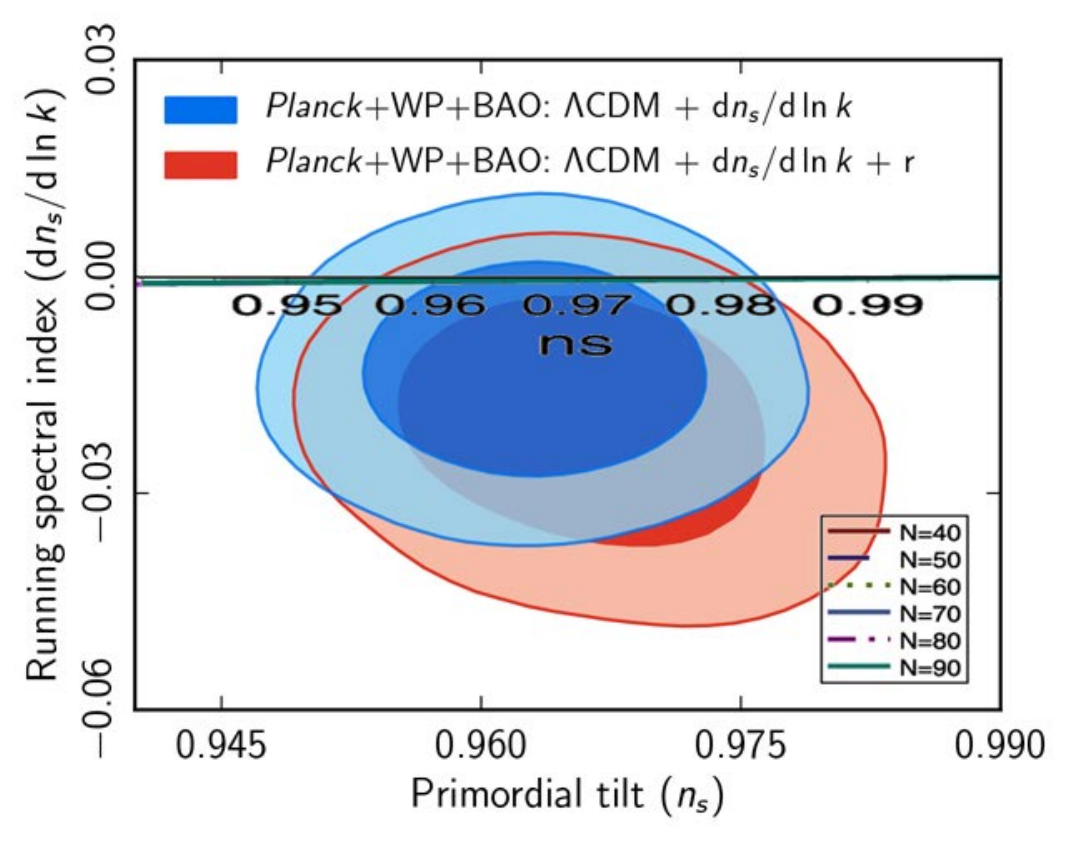}
\caption{\label{fig:1}\small{Plots of tensor to scalar ratio
(left) and running of scalar spectral index (right)
versus scalar spectral index, for a non-minimally kinetic coupled DGP
model with $\zeta=10^{-6}$ and the potential $V(\phi)={\sigma}\phi^{\frac{1}{2}}$. The plots are depicted for six values
of number of e-folding $N$. For all given viable values of $N$, the
non-minimally kinetic coupled DGP model lies inside the $95\%$ CL of the
Planck+WP+highL data, but it does not lie in the $95\%$ CL of the Planck+WP+highL+BICEP2
data. The values of  running of  spectral index are very
close to zero for all given values of $N$, and coincides  with the prediction
of the single monomial chaotic inflationary models with $50 < N <
60$ (For comparison, there is a   purple strip hidden behind the running
spectral indexes of non-minimally kinetic coupled DGP model, showing the running
spectral index of single monomial chaotic inflationary models with $50 < N <
60$). The non-minimally kinetic coupled DGP model lies inside the $95\%$ CL of the
Planck+WP+BAO:$\Lambda$CDM $dn_s/dln_k$ data, but it does not lie in the $95\%$ CL of the Planck+WP+BAO:$\Lambda$CDM $dn_s/dln_k+r$ 
data. }}
\end{figure}

 The non-minimally kinetic coupled DGP
model is well inside the joint $95\%$ CL Planck+WP+highL data (red area)
for all values of e-folding $N$, but it does not  lie inside the joint $95\%$ CL Planck+WP+highL+BICEP2 data (blue area). In the left plot of  Fig.\ref{fig:1} the behavior of tensor to scalar ratio versus scalar
spectral index is shown in the background of the Planck+WP+highL+BICEP2  data  for six values of $N$. In the right plot,  the evolution of  running of spectral
index versus scalar spectral index has been plotted for the similar
situation. It is seen
that, for all six values of the number of e-folding, the running of
scalar spectral index is  negative and close to zero.
\subsection{$V(\phi)={\sigma}\phi^{\frac{2}{3}}$}
\begin{figure}
\includegraphics[width=8.55 cm]{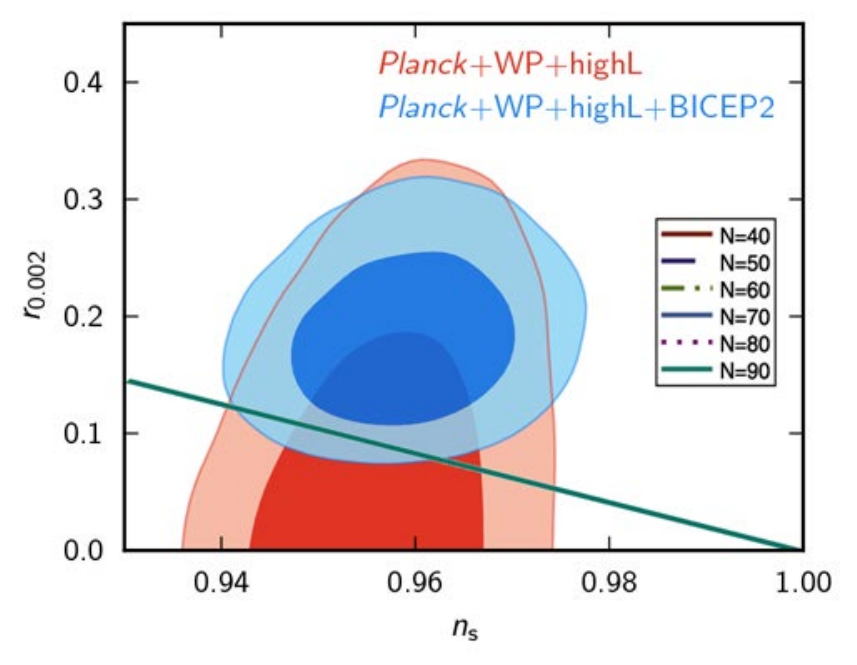}
\includegraphics[width=8.8 cm]{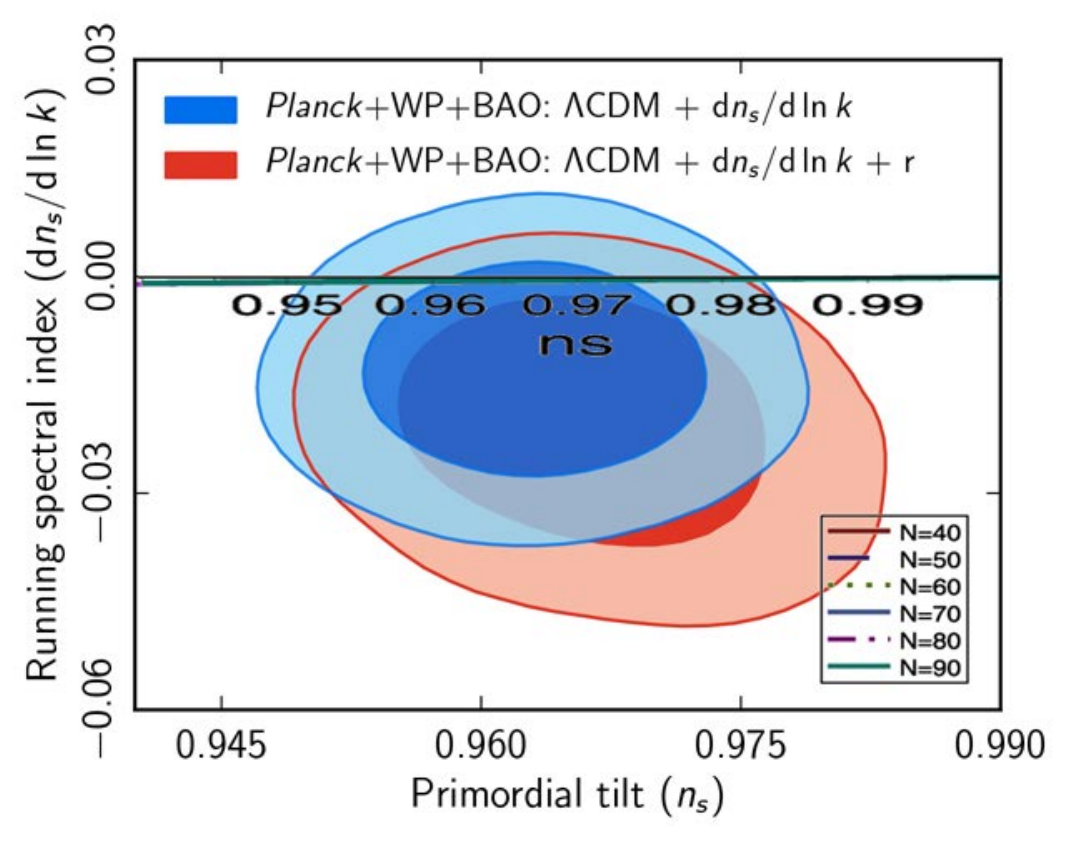}
\caption{\label{fig:2}\small{Plots of tensor to scalar ratio
(left) and running of scalar spectral index (right)
versus scalar spectral index, for a non-minimally kinetic coupled DGP
model with $\zeta=10^{-6}$ and $V(\phi)={\sigma}\phi^{{2/3}}$ potential. The figure has been plotted for six values of the number of e-folding. For all given viable values of $N$, the
non-minimally kinetic coupled DGP model is well inside the $95\%$ CL Planck+WP+highL
data, but it does not lie within the $95\%$ CL Planck+WP+highL+BICEP2 data. The values of running of spectral index are
very close to zero for all given values of $N$. The non-minimally kinetic coupled DGP model lies inside the $95\%$ CL of the
Planck+WP+BAO:$\Lambda$CDM $dn_s/dln_k$ data, but it does not lie in the $95\%$ CL of the Planck+WP+BAO:$\Lambda$CDM $dn_s/dln_k+r$ 
data. }}
\end{figure}

The non-minimally kinetic coupled DGP
model is well inside the $95\%$ CL of the Planck+WP+highL data, but it
does not lie inside the $95\%$ CL of the Planck+WP+highL+BICEP2 data. Evolution of tensor to scalar
ratio versus  scalar spectral index is shown in the left plot of
Fig.\ref{fig:2}. For all given values of $N$, the non-minimally kinetic coupled DGP model
 lies inside the $95\%$ CL Planck+WP+highL data. Evolution
of the running of  scalar spectral index versus scalar
spectral index has been plotted in the right panel of Fig.\ref{fig:2}. For this case,
the running is negative and close to zero.

\subsection{$V(\phi)={\sigma}\phi$}
\begin{figure}
\includegraphics[width=8.55 cm]{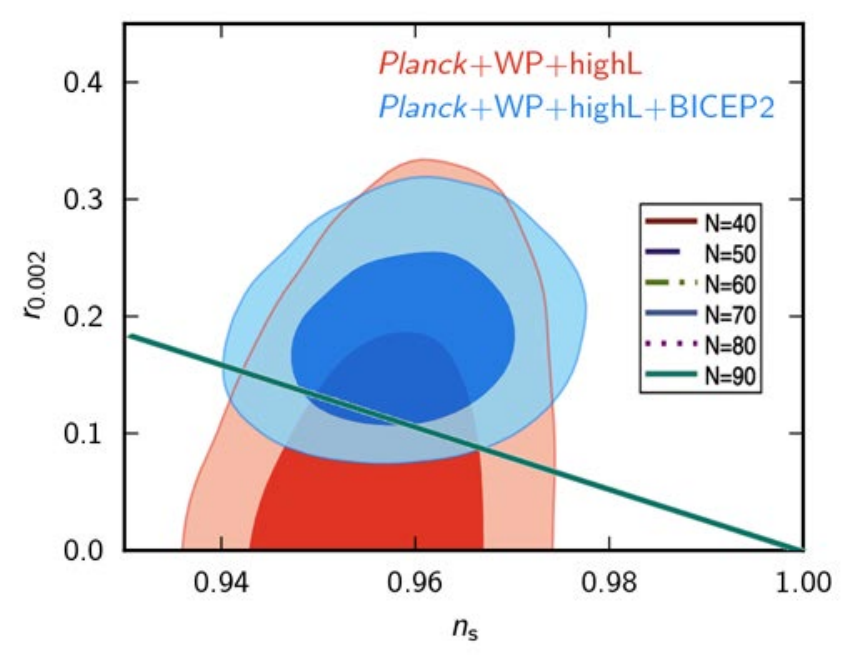}
\includegraphics[width=8.8 cm]{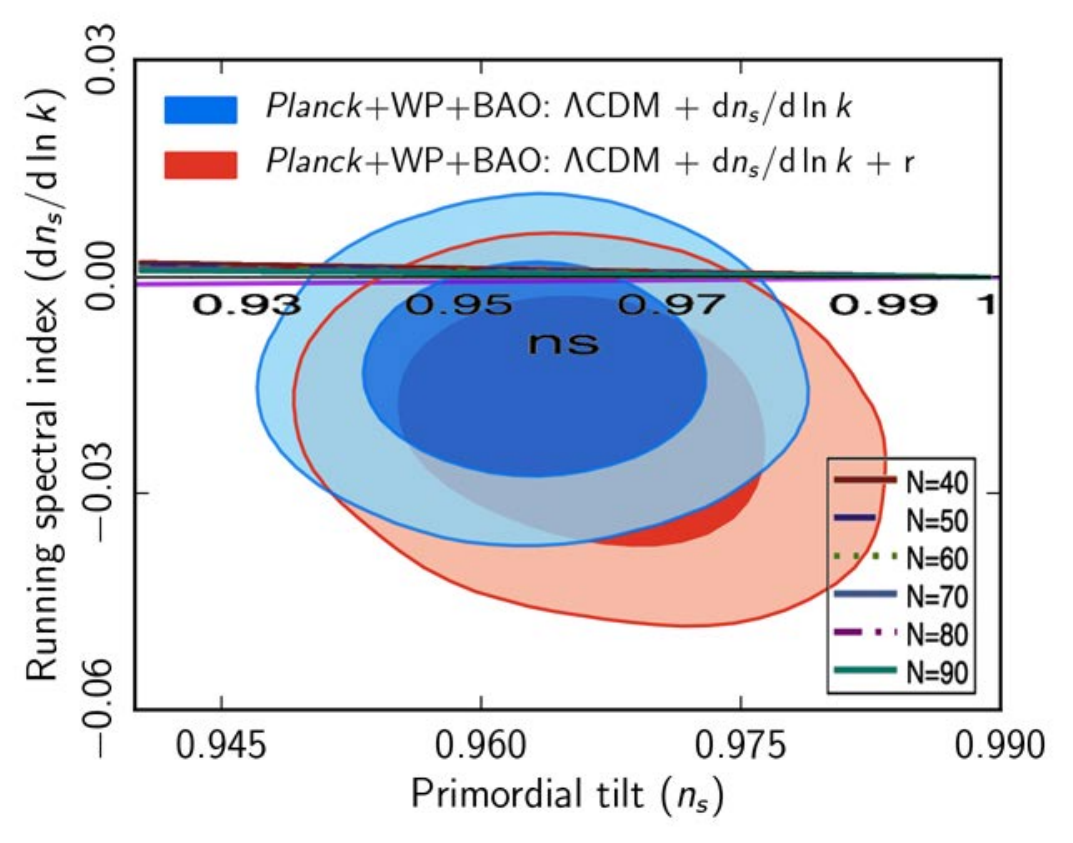}
\caption{\label{fig:3}\small{Plots of tensor to scalar ratio
(left) and running of scalar spectral index (right)
versus scalar spectral index, for a non-minimally kinetic coupled DGP
model with $\zeta=10^{-6}$ and a linear potential
$V(\phi)\sim\phi$. The figure has been plotted for six values
of number of e-folding. For all given viable values of $N$, the
non-minimally kinetic coupled DGP model is well inside the $95\%$ CL Planck+WP+highL data and is almost inside the $95\%$ CL Planck+WP+highL+BICEP2 data. The values of running of scalar spectral index are
almost zero for all given values of $N$. }The non-minimally kinetic coupled DGP model lies inside the $95\%$ CL of the
Planck+WP+BAO:$\Lambda$CDM $dn_s/dln_k$ data, but it does not lie in the $95\%$ CL of the Planck+WP+BAO:$\Lambda$CDM $dn_s/dln_k+r$ 
data.}
\end{figure}

 A minimally coupled four-dimensional setup with this potential lies
within the $95\%$ CL of the Planck+WMAP9+BAO data [26]. Our
braneworld model (non-minimally kinetic coupled model),
with this linear potential, lies within the $95\%$ CL Planck+WP+highL+BICEP2. A minimally coupled DGP model with this potential
lies still inside the $95\%$ CL of the Planck+WMAP9+BAO data. As before, we consider six values of number of e-folding.
In the left plot of Fig.\ref{fig:3}, we see the evolution of tensor to scalar ratio versus scalar spectral index. From our numerical analysis it appears that in a DGP model with
non-minimally kinetic coupled gravity, the model lies in the $95\%$ CL Planck+WP+highL+BICEP2,
for all given values of $N$. The
right plot of Fig.\ref{fig:3} shows the evolution of  running
of scalar spectral index versus scalar spectral index. As
the figure shows, for a non-minimally kinetic coupled DGP model with a
linear potential, the running of scalar spectral index is
close to zero.

\subsection{$V(\phi)={\sigma}\phi^{2}$}
\begin{figure}
\includegraphics[width=8.55 cm]{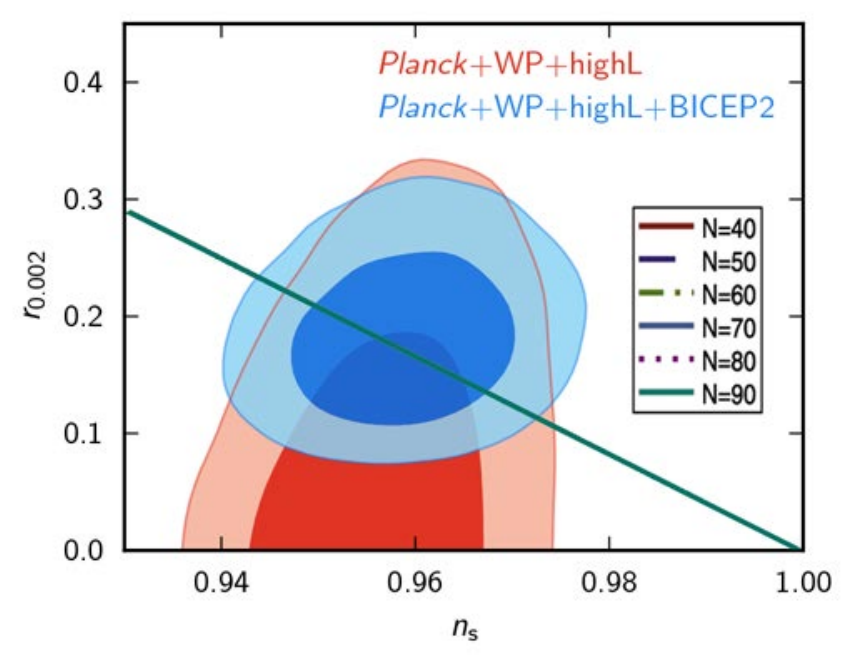}
\includegraphics[width=8.8 cm]{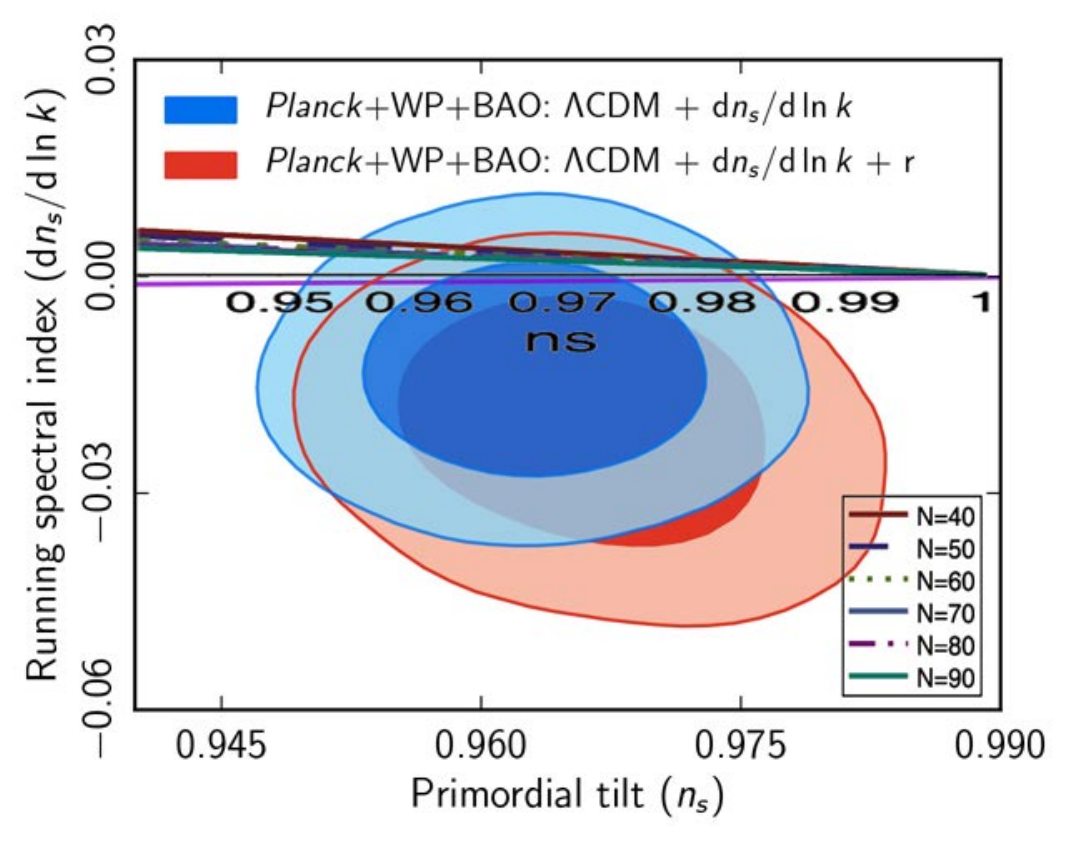}
\caption{\label{fig:4}\small{Plots of tensor to scalar ratio
(left) and running of scalar spectral index (right)
versus scalar spectral index, for a non-minimally kinetic coupled DGP
model   with $\zeta=10^{-6}$ and a quadratic potential $V(\phi)\sim\phi^{2}$. The figure has been plotted for six values
of number of e-folding. For all given viable values of $N$, the
non-minimally kinetic coupled DGP model is well inside the
$95\%$ CL Planck+WP+highL+BICEP2 data and is almost inside the $95\%$ CL Planck+WP+highL data. The values of the running of spectral index are almost close to zero for all given values of $N$. }The non-minimally kinetic coupled DGP model lies almost inside the $95\%$ CL of the
Planck+WP+BAO:$\Lambda$CDM $dn_s/dln_k$ data, but it does not lie in the $95\%$ CL of the Planck+WP+BAO:$\Lambda$CDM $dn_s/dln_k+r$ 
data.}
\end{figure}

In
\cite{26}, it has been shown that in 4-dimensions the model with this
potential lies outside and inside the $95\%$ CL of the joint Planck+WMAP9+BAO
data for $N=50$ and $N=60$, respectively. Now, we explore the
situation for a 5-dimensional model.
According to the WMAP7+BAO+H$_{0}$ data \cite{30}, a warped DGP model
with minimally coupled scalar field and with a squared potential,
lies inside the $95\%$ CL for $N< 70$. Now, with recent BICEP2 date, the situations change
considerably. In a minimally coupled DGP model with a quadratic
potential, for all $N\geq 40$, the model is outside the joint $95\%$
CL of the Planck+WMAP9+BAO data. In our model for a non-minimally kinetic
coupled DGP
model, for all given values of $N$ the model is well inside the joint $95\%$ CL Planck+WP+highL+BICEP2 data. The left plot of Fig.\ref{fig:4} shows
the behavior of tensor to scalar ratio versus scalar
spectral index in the background of the Planck+WP+highL+BICEP2 data. This
figure has been plotted for six values of $N$. Also, we have plotted the evolution of running of scalar spectral
index versus scalar spectral index in the background of the
 Planck+WP+highL+BICEP2 data (the left panel of Fig.4). We see
that, for all six values of the number of e-folding, the running of
scalar spectral index is  close to zero.
\subsection{$V(\phi)={\sigma}\phi^{3}$}
\begin{figure}
\includegraphics[width=8.55 cm]{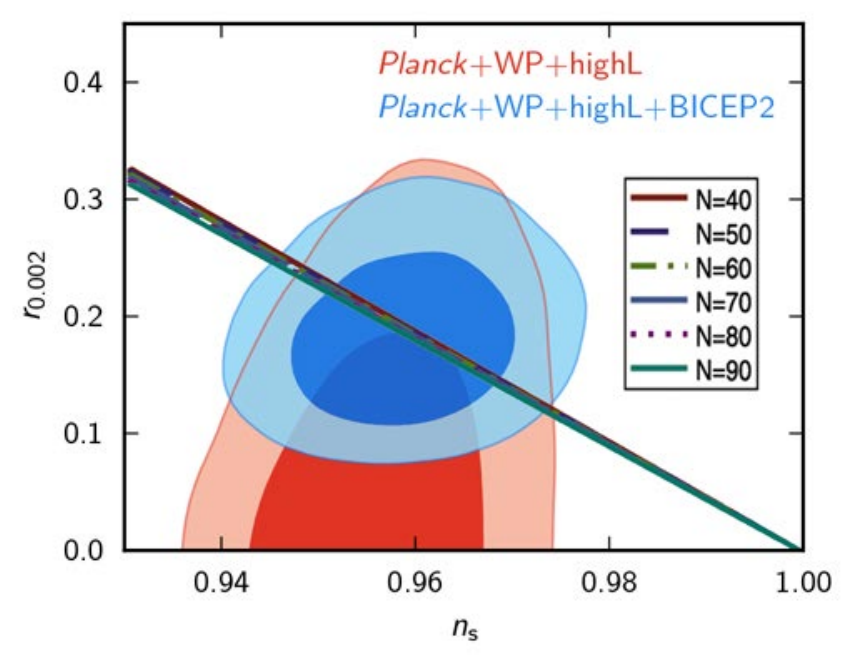}
\includegraphics[width=8.8 cm]{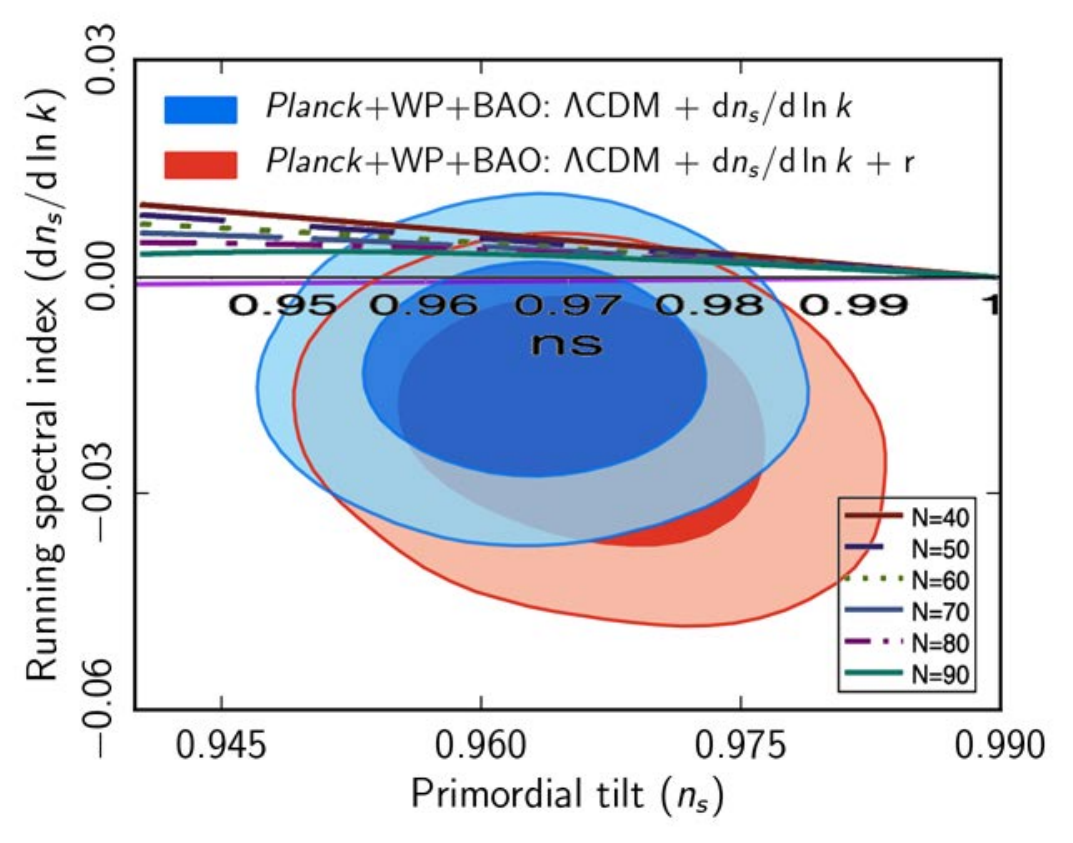}
\caption{\label{fig:5}\small{Plots of tensor to scalar ratio
(left) and running of scalar spectral index (right)
versus scalar spectral index, for a non-minimally kinetic coupled DGP
model with $\zeta=10^{-6}$ and a cube potential $V(\phi)\sim\phi^{3}$. The figure has been plotted for
$N=40,\,50,\,60,\,70,\,80\,$ and $90$.  For all given viable values of $N$ the non-minimally kinetic coupled DGP model is well inside the $95\%$ CL Planck+WP+highL+BICEP2 data and is almost inside the $95\%$ CL Planck+WP+highL data. The values of  running of spectral index are
approximately zero for all given values of $N$. }The non-minimally kinetic coupled DGP model lies approximately close to the $95\%$ CL of the
Planck+WP+BAO:$\Lambda$CDM $dn_s/dln_k$ data, but it does not lie in the $95\%$ CL of the Planck+WP+BAO:$\Lambda$CDM $dn_s/dln_k+r$ 
data.}
\end{figure}

 It has been shown in \cite {QingGao} that in natural inflation this potential lies in the $95\%$ CL Planck+WP+highL+BICEP2 and also
it has been
confirmed with WMAP9 \cite{31} and Planck \cite{26} data that a model with
a cube potentia in 4-dimensions lies outside the $95\%$ CL. In our
branworld model, we obtain a different result: a non-minimally kinetic coupled DGP
model with this potential, lies inside the $95\%$ CL Planck+WP+highL+BICEP2 data for each given value of $N$. The results are shown in Fig.\ref{fig:5}.
 Note
that the evolution of running of scalar spectral index
corresponding to the cube potential is shown in the right plot
of Fig.\ref{fig:5}. The value of running of scalar spectral
index is close to zero.

\subsection{$V(\phi)=\sigma\phi^4$}
 A
minimally coupled 4-dimensional model with this potential lies
within the $95\%$ CL of the Planck+WMAP9+BAO data \cite{26}. For $N \leq70$, our braneworld model (non-minimal kinetic coupled)
with this  potential lies inside the $95\%$ CL Planck+WP+highL+BICEP2 data. But for all given values
 of $N$ the model
 lies inside the $95\%$ CL Planck+WP+highL data. The running of spectral
 index is plotted in the right plot of Fig.6.
In the left plot of Fig.6, we see the evolution of tensor to scalar ratio versus scalar spectral index.

\begin{figure}[htp]
\includegraphics[width=8.55 cm]{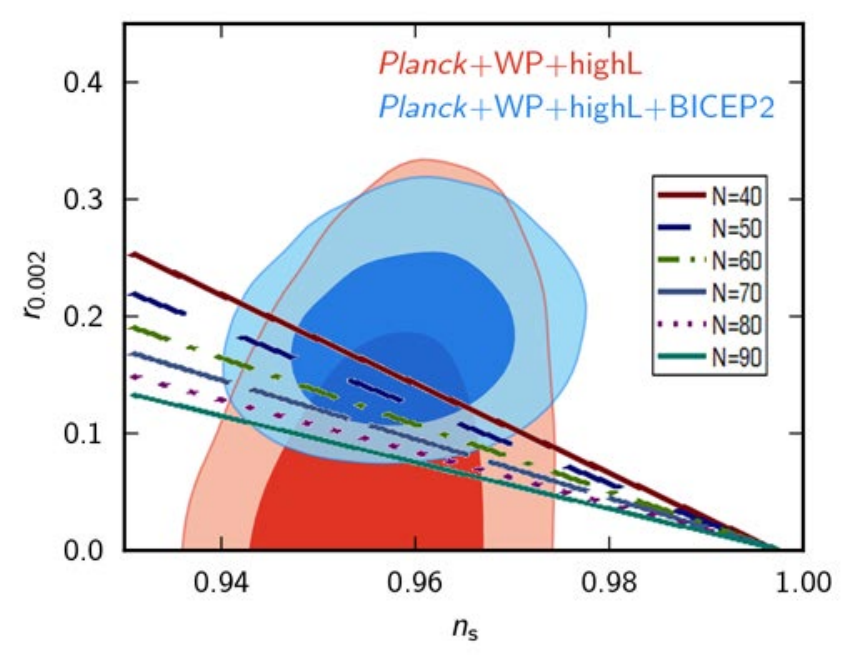}
\includegraphics[width=8.8 cm]{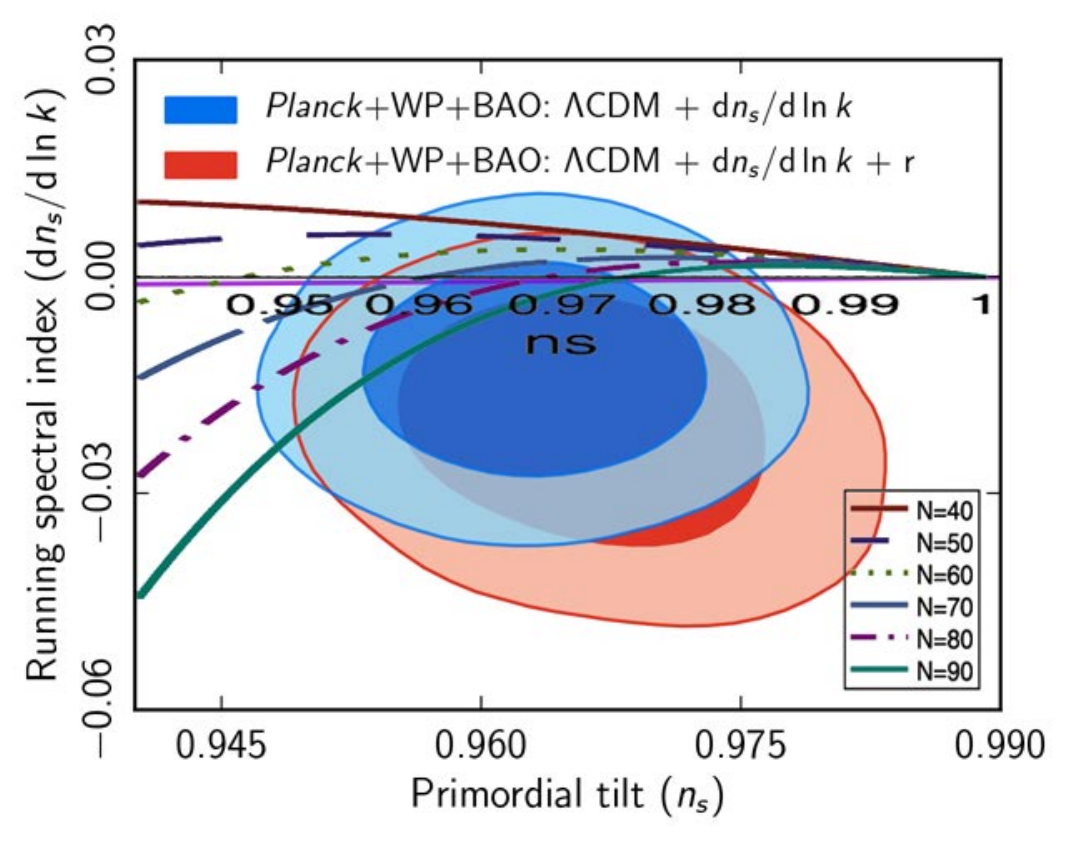}
\caption{\label{fig:6}\small{Plots of tensor to scalar ratio
(left) and running of scalar spectral index (right)
versus scalar spectral index, for a non-minimally kinetic coupled DGP
model with $\zeta=10^{-6}$ and a potential of type
$V(\phi)\propto\phi^{4}$. The figure has been plotted for
$N=40,\,50,\,60\,70,\,80,\,$ and $90$. For all given values of $N$  
 the model
 lies inside the $95\%$ CL Planck+WP+highL data and
 lies inside the $95\%$ CL Planck+WP+highL+BICEP2 data just for $N=40,\,50,\,60,\,70$. The running of spectral
 index is plotted in the right panel and it is close to zero. }Neither
of $N=40,\,50,\,60,\,70$ lies inside the $95\%$ CL of the
Planck+WP+BAO:$\Lambda$CDM $dn_s/dln_k$ data and the Planck+WP+BAO:$\Lambda$CDM $dn_s/dln_k+r$ 
data.}
\end{figure}

\subsection{$V(\phi)=\sigma\phi^5$}
Similar to the other cases, we consider six values of the number of e-folding.
In the left plot of Fig.7, we see the evolution of tensor to scalar ratio versus scalar spectral index.
For all given values of $N$, the non-minimally kinetic coupled DGP braneworld model with this  potential lies inside the $95\%$ CL Planck+WP+highL data, but does not lie inside the $95\%$ CL Planck+WP+highL+BICEP2 . The running of spectral index is plotted in the right plot of Fig.7 and it is close to zero.
\begin{figure}[htp]
\includegraphics[width=8.55 cm]{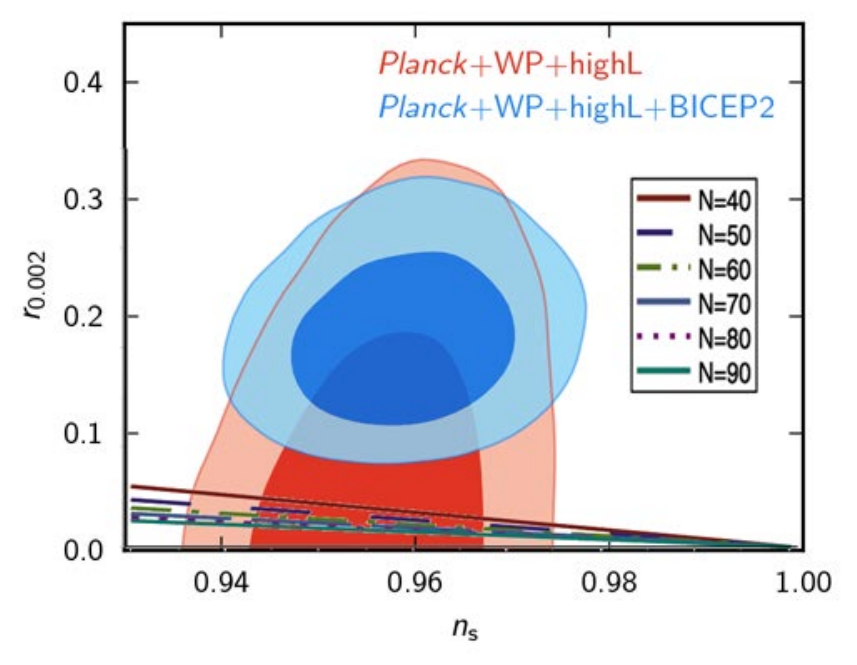}
\includegraphics[width=8.8 cm]{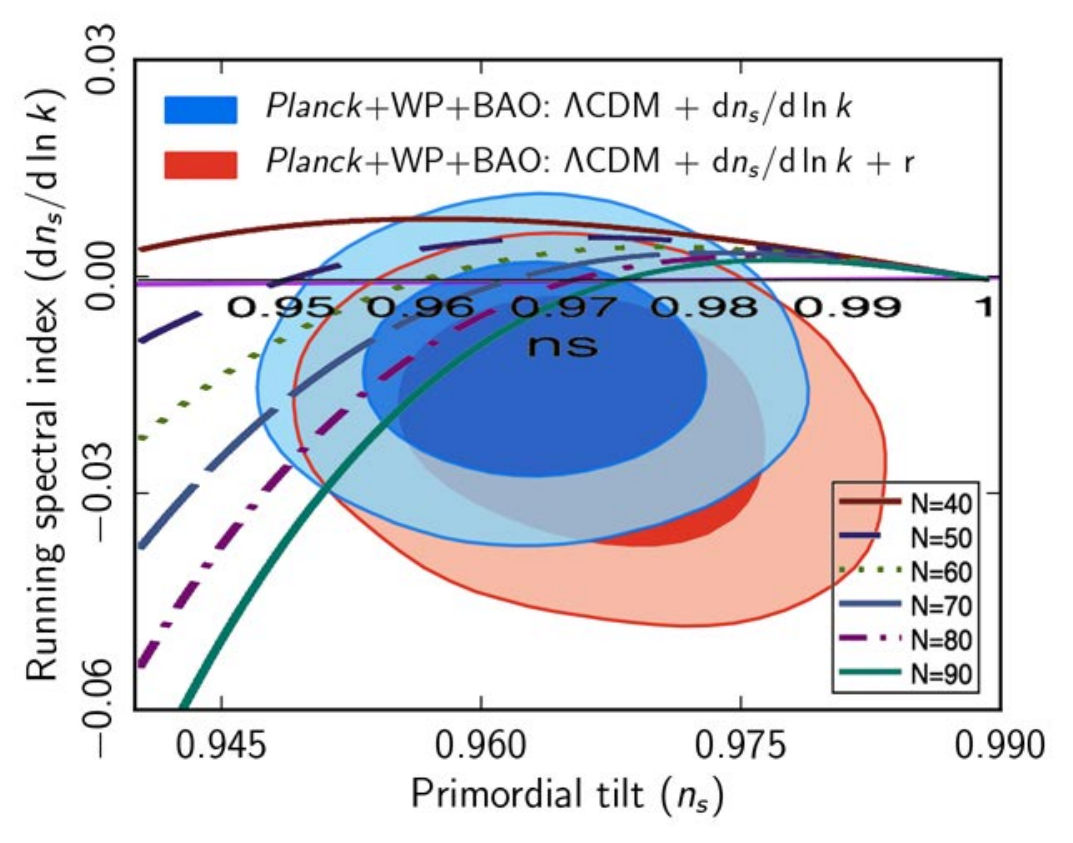}
\caption{\label{fig:7}\small{Plots of tensor to scalar ratio
(left) and running of scalar spectral index (right)
versus scalar spectral index, for a non-minimally kinetic coupled DGP
model with $\zeta=10^{-6}$ and a potential of the type
$V(\phi)\sim\phi^{5}$. The figure has been plotted for
$N=40,\,50,\,60,\,70,\,80,$ and $90$. For all given
viable values of $N$, the non-minimally kinetic coupled DGP model
lies inside the $95\%$ CL Planck+WP+highL data, but it does not lie inside the $95\%$ CL Planck+WP+highL+BICEP2 data. The running of spectral index is plotted in the right plot of the figure;  it is negative for $N\geq50$, and is positive (close to zero) for $N=40$. }The model  with values of $N=70,\,80,\,90\,$ and $N=\,80,\,90\,$ lie inside the $95\%$ CL of the Planck+WP+BAO:$\Lambda$CDM $dn_s/dln_k$ data and the Planck+WP+BAO:$\Lambda$CDM $dn_s/dln_k+r$ data, respectively.}
\end{figure}

Our numerical analysis for a DGP model with a non-minimally kinetic coupled provides us with the value of each of the parameters $\phi_i $ (the value of $\phi$ at the beginning of inflation), $\phi_e$ (the value of $\phi$ at the end of inflation), $ H_i$ (the value of $H$ at the beginning of inflation), $H_e$ (the value of $H$ at the end of inflation), $\varepsilon,  \eta$ (the slow-roll parameters), $V(\phi_i)$ (the value of potential at the beginning of inflation), and $9\zeta H_i^2 {\dot{\phi_i}}^2 $(to control the slow-roll condition $9\zeta H^2\dot{\phi}^{2}\ll V(\phi))$  for $N=60$ (the number of e-folding), in the following tables.
 These values can tell us ``which potentials are in best agreement with the recent observations."

It turns out  that the potentials  $V(\phi)\propto \phi^2$ and $V(\phi)\propto \phi^3$ provide respectively the best fits with the observations (see Fig.4, Fig.5 and table.II).

\begin{table*}
\begin{tiny}
\caption{\label{tab:1}  }
\begin{tabular}{cccccccccccc}
\\ \hline \hline $V(\phi)$&$\sigma$&$\zeta$&$N$&$\phi_e$&$H_e$&$\phi_i$&$H_i$&$\varepsilon$&$\eta$&$9f
H_i^2 {\dot{\phi_i}}^2$&$V(\phi_i)$\\ \hline\\
${\sigma}\phi^{\frac{1}{2}}$&$1$&$0.001$&$60$&$0.3535$& $0.25705006\sqrt{3}$& $7.76$ &$0.5564\sqrt{3}$&$0.0020$&$-0.0062$&$0.007755$&$2.7845$\\\\
${\sigma}\phi^{\frac{2}{3}}$&$1$&$0.001$&$60$&$0.4113$& $0.25943584\sqrt{3}$& $8.942$ &$0.6919\sqrt{3}$&$0.0027$&$-0.0054$&$0.01856$&$4.308$\\\\
${\sigma}\phi$&$1$&$0.001$&$60$&$0.7069$& $0.28030475\sqrt{3}$& $10.977$ &$1.1024\sqrt{3}$&$0.0041$&$-0.0040$&$0.11967$&$10.939$\\\\
${\sigma}\phi^{2}$&$1$&$0.001$&$60$&$1.4136$& $0.47123562\sqrt{3}$& $14.782$ &$4.9273\sqrt{3}$&$0.0075$&$0.000001$&$47.7464$&$218.5075$\\\\
${\sigma}\phi^{3}$&$1$&$0.001$&$60$&$0.00008$& $0.00004803\sqrt{3}$& $13.514$ &$16.5601\sqrt{3}$&$0.0071$&$0.00098$&$6091.7848$&$2468.1470$\\\\
${\sigma}\phi^{4}$&$1$&$0.001$&$60$&$0.0017$& $0.00004805\sqrt{3}$& $11.047$ &$40.6787\sqrt{3}$&$0.0041$&$0.0007$&$2.218\times10^5$&$14892.8363$\\\\
${\sigma}\phi^5$&$1$&$0.001$&$60$&$0.0080$& $0.00004811\sqrt{3}$& $7.902$ &$58.5089\sqrt{3}$&$0.0063$&$0.0012$&$9.49\times
10^6$&$30809.5338$\\\\
${\sigma}\phi^6$&$1$&$0.001$&$60$&$0.0203$& $0.00004820\sqrt{3}$& $6.731$ &$101.6523\sqrt{3}$&$0.0042$&$0.0094$&$8.46\times
10^6$&$92998.8$\\\\
${\sigma}e^{\phi}$&$1$&$0.001$&$-$&$"Imaginary"$& $-$& $-$ &$-$&$-$&$-$&$-$&$-$\\\\
\end{tabular}
\end{tiny}
\end{table*}

\begin{figure}[htp]
\includegraphics[width=8.55 cm]{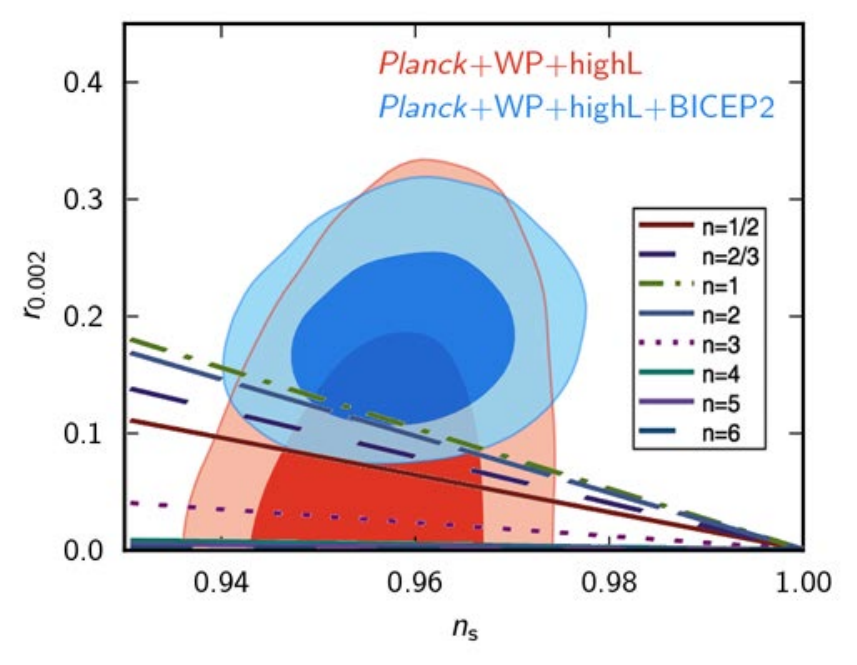}
\includegraphics[width=8.8 cm]{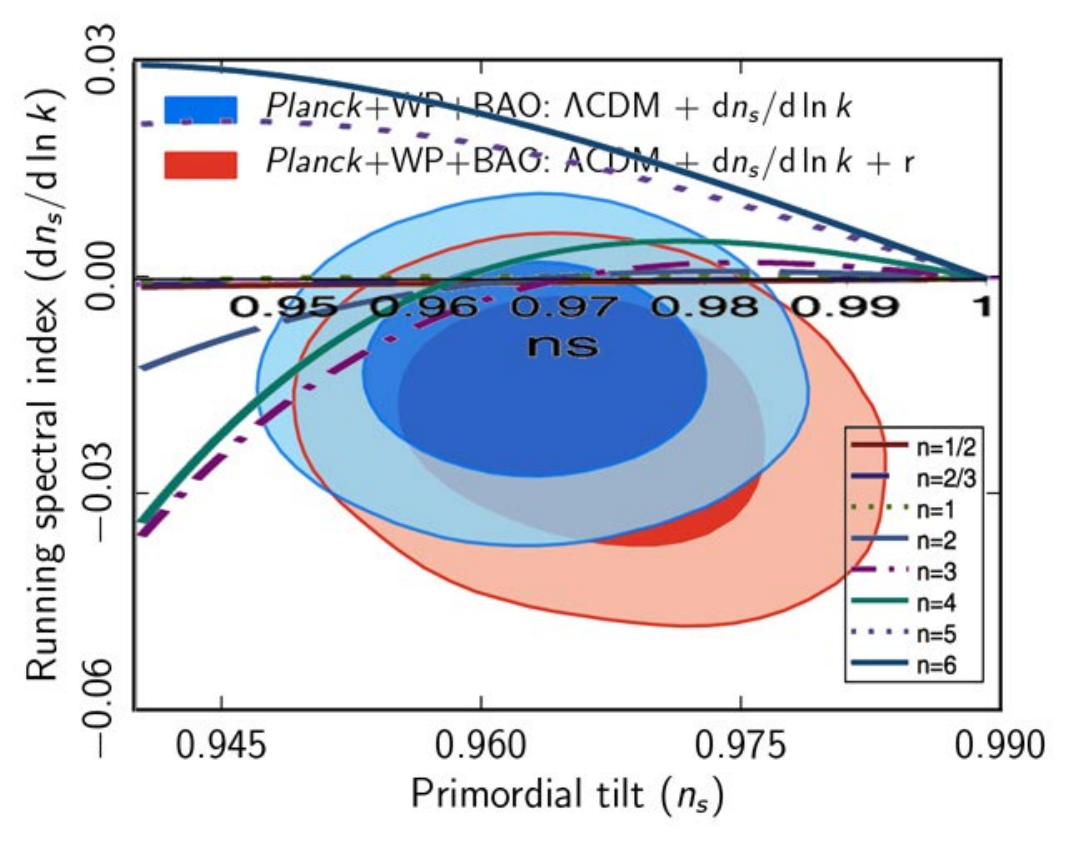}
\caption{\label{fig:8}\small{{ Plots of tensor to scalar ratio
(left) and running of scalar  spectral index (right)
versus scalar spectral index, for a non-minimally kinetic coupled DGP
model with $\zeta=10^{-3}$ and a potential of the type
$V(\phi)\sim\phi^{n}$. The figure has been plotted for
$N=60$. For all given
values of $n$, the non-minimally kinetic coupled DGP model
lies inside the $95\%$ CL Planck+WP+highL data, but it does not lie inside the $95\%$ CL Planck+WP+highL+BICEP2 data, except  $n=1, 2$. The running of spectral index is plotted in the right plot of the figure.  Neither
of the values of $n$ lies inside the $95\%$ CL of the
Planck+WP+BAO:$\Lambda$CDM $dn_s/dln_k+r$ data and all the values of $n$
except $n=5, 6$ lie inside the $95\%$ CL of the Planck+WP+BAO:$\Lambda$CDM $dn_s/dln_k$ data.
}}}
\end{figure}
\begin{figure}[htp]
\includegraphics[width=8.55 cm]{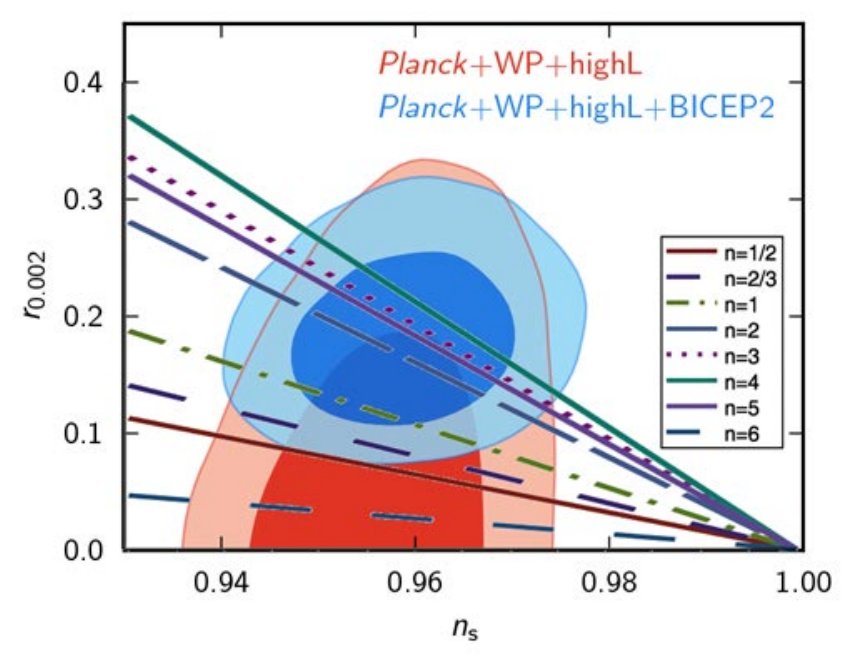}
\includegraphics[width=8.8 cm]{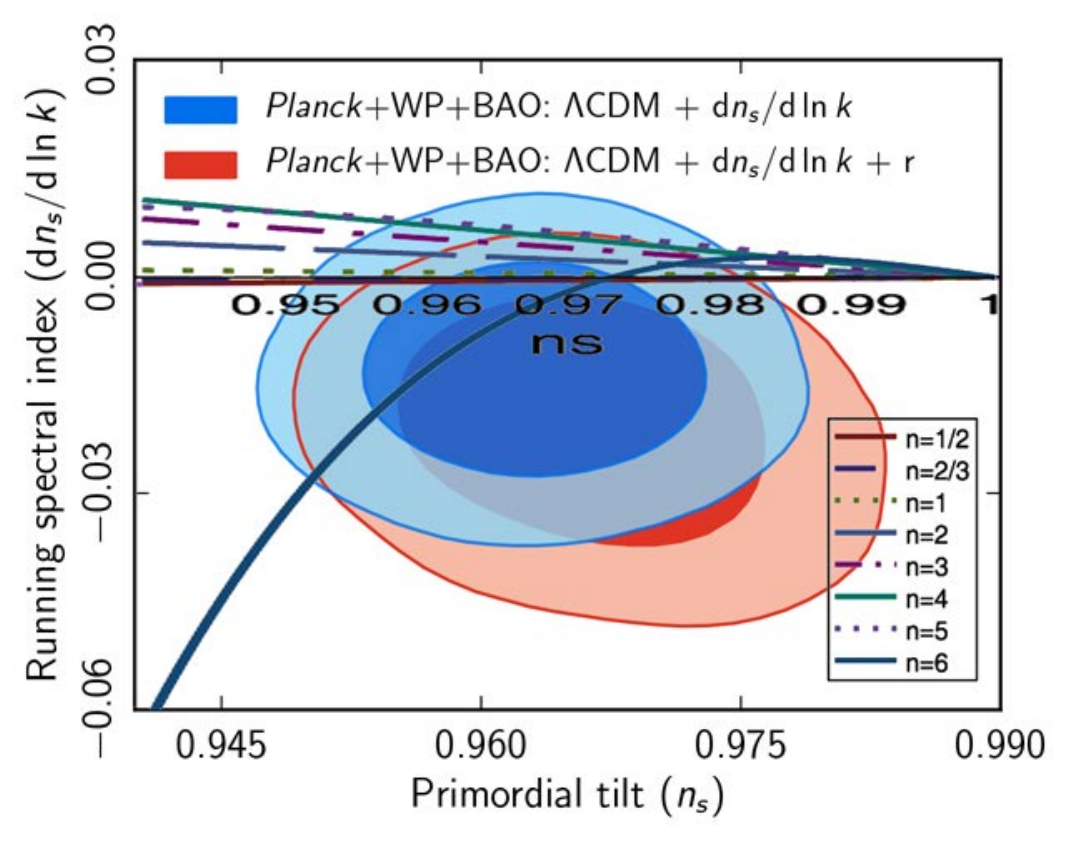}
\caption{\label{fig:9}\small{{ Plots of tensor to scalar ratio
(left) and running of scalar  spectral index (right)
versus scalar spectral index, for a non-minimally kinetic coupled DGP
model with $\zeta=10^{-8}$ and a potential of the type
$V(\phi)\sim\phi^{n}$. The figure has been plotted for
$N=60$. For all given
values of $n$, the non-minimally kinetic coupled DGP model
lies inside the $95\%$ CL Planck+WP+highL data except  $n=3, 4$, and lies inside the $95\%$ CL Planck+WP+highL+BICEP2 data, except for $n=1/2, 2/3,
6$. The running of spectral index is plotted in the right plot of the figure. Neither of the values of $n$ lies inside the $95\%$ CL of the
Planck+WP+BAO:$\Lambda$CDM $dn_s/dln_k+r$ data except $n=6$, and just the values  $n=1, 2/3, 1/2$ lie inside the $95\%$ CL of the Planck+WP+BAO:$\Lambda$CDM $dn_s/dln_k$ data.}}}
\end{figure}
\begin{figure}[htp]
\includegraphics[width=8.55 cm]{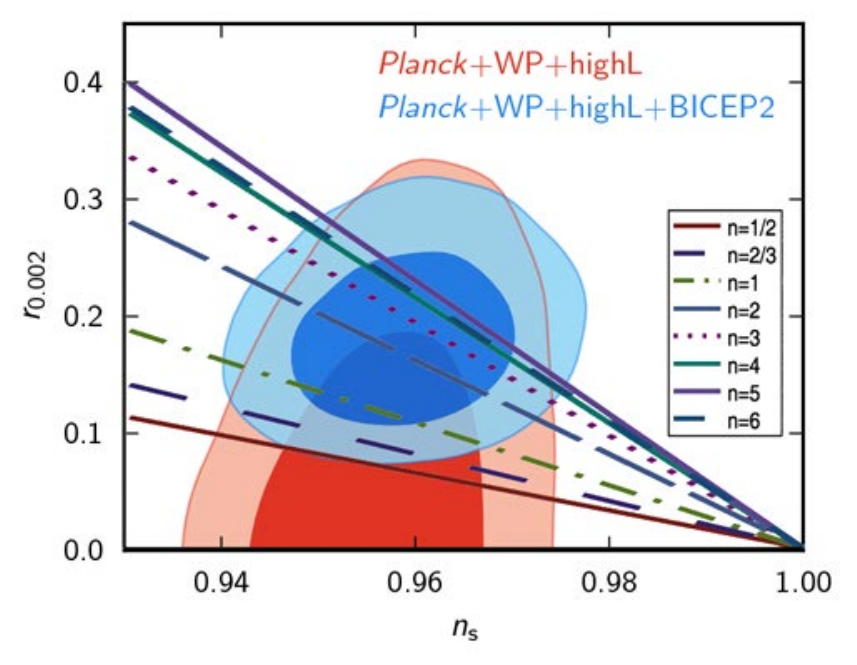}
\includegraphics[width=8.8 cm]{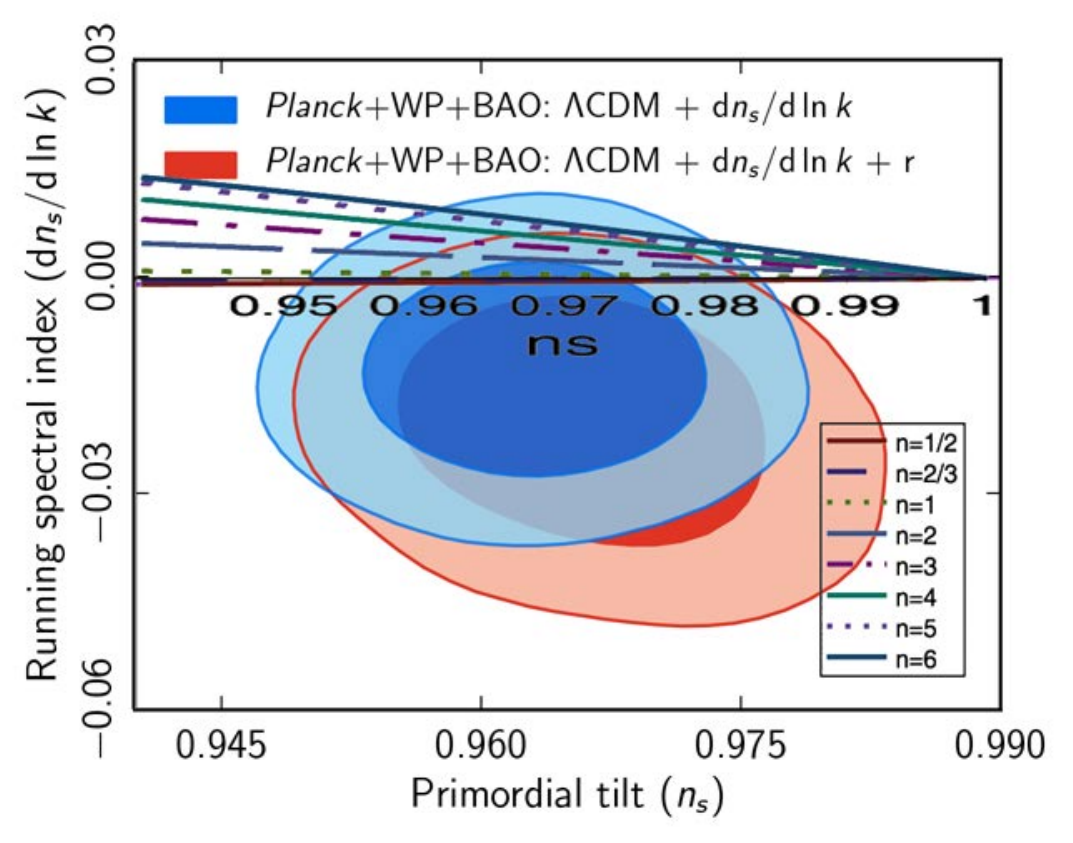}
\caption{\label{fig:10}\small{{Plots of tensor to scalar ratio
(left) and running of scalar  spectral index (right)
versus scalar spectral index, for a non-minimally kinetic coupled DGP
model with $\zeta=10^{-10}$ and a potential of the type
$V(\phi)\sim\phi^{n}$. The figure has been plotted for
$N=60$. For all given
values of $n$, the non-minimally kinetic coupled DGP model
lies inside the $95\%$ CL Planck+WP+highL data except  $n=3, 4, 5, 6$, and lies inside the $95\%$ CL Planck+WP+highL+BICEP2 data, except  $n=1/2, 2/3$. The running of spectral
 index is plotted in the right plot of the figure. Neither
of the values of $n$ lies inside the $95\%$ CL of the
Planck+WP+BAO:$\Lambda$CDM $dn_s/dln_k+r$ data, and just the values  $n=1, 2/3, 1/2$ lie inside the $95\%$ CL of the Planck+WP+BAO:$\Lambda$CDM $dn_s/dln_k$ data.}}}
\end{figure}

\begin{table*}
\begin{tiny}
\caption{\label{tab:1}  }
\begin{tabular}{cccccccccccc}
\\ \hline \hline $V(\phi)$&$\sigma$&$\zeta$&$N$&$\phi_e$&$H_e$&$\phi_i$&$H_i$&$\varepsilon$&$\eta$&$9f
H_i^2 {\dot{\phi_i}}^2$&$V(\phi_i)$\\ \hline\\
${\sigma}\phi^{\frac{1}{2}}$&$1$&$0.000001$&$60$&$0.3535$& $0.25705006\sqrt{3}$& $7.7540$ &$0.5562\sqrt{3}$&$0.0020$&$-0.0062$&$0.000013$&$2.7548$\\\\
${\sigma}\phi^{\frac{2}{3}}$&$1$&$0.000001$&$60$&$0.4113$& $0.25943584\sqrt{3}$& $8.9570$ &$0.6922\sqrt{3}$&$0.0027$&$-0.0055$&$0.00003$&$4.3130$\\\\
${\sigma}\phi$&$1$&$0.000001$&$60$&$0.7070$& $0.28030475\sqrt{3}$& $10.977$ &$1.1044\sqrt{3}$&$0.0041$&$-0.0041$&$0.00020$&$10.9770$\\\\
${\sigma}\phi^{2}$&$1$&$0.000001$&$60$&$1.4141$& $0.00004803\sqrt{3}$& $15.560$ &$5.1867\sqrt{3}$&$0.0082$&$2.5938\times10^{-8}$&$0.09862$&$242.1146$\\\\
${\sigma}\phi^{3}$&$1$&$0.000001$&$60$&$0.00008$& $0.00004803\sqrt{3}$& $18.932$ &$27.4583\sqrt{3}$&$0.0124$&$0.0040$&$77.4665$&$6785.6192$\\\\
${\sigma}\phi^{4}$&$1$&$0.000001$&$60$&$0.0017$& $0.00004805\sqrt{3}$& $20.850$ &$144.9075\sqrt{3}$&$0.0140$&$0.0047$&$60087.2883$&$188983.6520$\\\\
${\sigma}\phi^5$&$1$&$0.000001$&$60$&$0.0080$& $0.00004811\sqrt{3}$& $17.908$ &$452.4043\sqrt{3}$&$0.0095$&$0.0022$&$5.7\times
10^6$&$1.84\times 10^6$\\\\
${\sigma}\phi^6$&$1$&$0.000001$&$60$&$0.0203$& $0.00004820\sqrt{3}$& $13.760$ &$868.5117\sqrt{3}$&$0.0077$&$0.0018$&$7.7\times
10^7$&$6.8\times10^6$\\\\
${\sigma}e^{\phi}$&$1$&$0.000001$&$-$&$"Imaginary"$& $-$& $-$ &$-$&$-$&$-$&$-$&$-$\\\\
\end{tabular}
\end{tiny}
\end{table*}

\begin{table*}
\begin{tiny}
\caption{\label{tab:1}  }
\begin{tabular}{cccccccccccc}
\\ \hline \hline $V(\phi)$&$\sigma$&$\zeta$&$N$&$\phi_e$&$H_e$&$\phi_i$&$H_i$&$\varepsilon$&$\eta$&$9f
H_i^2 {\dot{\phi_i}}^2$&$V(\phi_i)$\\ \hline\\
${\sigma}\phi^{\frac{1}{2}}$&$1$&$10^{-8}$&$60$&$0.3535$& $0.25704370\sqrt{3}$& $7.76$ &$0.5564\sqrt{3}$&$0.0020$&$-0.0062$&$7.7\times10^{-10}$&$2.7856$\\\\
${\sigma}\phi^{\frac{2}{3}}$&$1$&$10^{-8}$&$60$&$0.4713$& $0.25943584\sqrt{3}$& $8.962$ &$0.6924\sqrt{3}$&$0.0027$&$-0.0055$&$1.86\times10^{-7}$&$4.3145$\\\\
${\sigma}\phi$&$1$&$10^{-8}$&$60$&$0.7070$& $0.28030475\sqrt{3}$& $10.977$ &$1.1045\sqrt{3}$&$0.0041$&$-0.0041$&$1.2\times10^{-6}$&$10.9770$\\\\
${\sigma}\phi^{2}$&$1$&$10^{-8}$&$60$&$1.4143$& $0.47146400\sqrt{3}$& $15.562$ &$5.1873\sqrt{3}$&$0.0082$&$1.9\times10^{-8}$&$0.00058$&$242.1758$\\\\
${\sigma}\phi^{3}$&$1$&$10^{-8}$&$60$&$0.00008$& $0.00004803\sqrt{3}$& $18.99$ &$27.5776\sqrt{3}$&$0.0124$&$0.0041$&$0.4685$&$6844.7143$\\\\
${\sigma}\phi^{4}$&$1$&$10^{-8}$&$60$&$0.0017$& $0.00004805\sqrt{3}$& $21.91$ &$160.016\sqrt{3}$&$0.0166$&$0.0082$&$531.0547$&$230446.1$\\\\
${\sigma}\phi^5$&$1$&$10^{-8}$&$60$&$0.0080$& $0.00004811\sqrt{3}$& $17.908$ &$962.8047\sqrt{3}$&$0.0197$&$0.0244$&$6.960\times
10^5$&$8.343\times 10^6$\\\\
${\sigma}\phi^6$&$1$&$10^{-8}$&$60$&$0.0203$& $0.00004820\sqrt{3}$& $25.437$ &$5486.2\sqrt{3}$&$0.0075$&$0.0020$&$7.33\times
10^8$&$2.7\times10^8$\\\\
${\sigma}e^{\phi}$&$1$&$10^{-8}$&$-$&$"Imaginary"$& $-$& $-$ &$-$&$-$&$-$&$-$&$-$\\\\
\end{tabular}
\end{tiny}
\end{table*}


\begin{table*}
\begin{tiny}
\caption{\label{tab:1}  }
\begin{tabular}{cccccccccccc}
\\ \hline \hline $V(\phi)$&$\sigma$&$\zeta$&$N$&$\phi_e$&$H_e$&$\phi_i$&$H_i$&$\varepsilon$&$\eta$&$9f
H_i^2 {\dot{\phi_i}}^2$&$V(\phi_i)$\\ \hline\\
${\sigma}\phi^{\frac{1}{2}}$&$1$&$10^{-10}$&$60$&$0.3535$& $0.25705007\sqrt{3}$& $7.7540$ &$0.5562\sqrt{3}$&$0.0020$&$-0.0062$&$7.7\times 10^{-10}$&$2.7845$\\\\
${\sigma}\phi^{\frac{2}{3}}$&$1$&$10^{-10}$&$60$&$0.4713$& $0.25943584\sqrt{3}$& $8.962$ &$0.6924\sqrt{3}$&$0.0027$&$-0.0055$&$1.86\times 10^{-9}$&$4.3145$\\\\
${\sigma}\phi$&$1$&$10^{-10}$&$60$&$0.7070$& $0.28030475\sqrt{3}$& $10.977$ &$1.1045\sqrt{3}$&$0.0041$&$-0.0041$&$1.2\times 10^{-8}$&$10.9770$\\\\
${\sigma}\phi^{2}$&$1$&$10^{-10}$&$60$&$1.4141$& $0.471392\sqrt{3}$& $15.560$ &$5.1867\sqrt{3}$&$0.0082$&$2.5\times10^{-8}$&$5.8\times10^{-6}$&$242.1745$\\\\
${\sigma}\phi^{3}$&$1$&$10^{-10}$&$60$&$0.00008$& $0.00004803\sqrt{3}$& $18.99$ &$27.5776\sqrt{3}$&$0.0124$&$0.0041$&$0.0047$&$6844.7143$\\\\
${\sigma}\phi^{4}$&$1$&$10^{-10}$&$60$&$0.0017$& $0.00004805\sqrt{3}$& $21.91$ &$160.0165\sqrt{3}$&$0.0140$&$0.0107$&$5.3105$&$230446.1$\\\\
${\sigma}\phi^5$&$1$&$10^{-10}$&$60$&$0.0080$& $0.00004811\sqrt{3}$& $17.908$ &$989.8567\sqrt{3}$&$0.0208$&$0.0022$&$7776.326$&$8.818\times 10^6$\\\\
${\sigma}\phi^6$&$1$&$10^{-10}$&$60$&$0.0203$& $0.00004820\sqrt{3}$& $26.72$ &$6358.9895\sqrt{3}$&$0.02043$&$0.0152$&$1.324\times
10^7$&$3.639\times10^8$\\\\
${\sigma}e^{\phi}$&$1$&$10^{-10}$&$-$&$"Imaginary"$& $-$& $-$ &$-$&$-$&$-$&$-$&$-$\\\\
\end{tabular}
\end{tiny}
\end{table*}
\section{conclusion and remarks}

In this paper, we have considered a $5D$ bulk spacetime together with a single $4D$ brane and derived the effective $4D$ gravitational equations. Then, we have studied the non-minimally kinetic coupled version of a braneworld gravity proposed by Dvali, Gabadadze, and Porrati, so called DGP model. We have derived the field equations, using the FRW metric accompanied by the perfect fluid, and studied the inflationary scenario in this model. Finally, we have confronted the numerical analysis of six typical  scalar field potentials with the observational data, and found that: \begin{itemize}
 
 \item For $V(\phi)=\sigma\phi^\frac{1}{2}$ and $V(\phi)=\sigma\phi^\frac{2}{3}$ and the given values of $N$, the non-minimally kinetic
coupled DGP model is well inside the $95\%$ CL of the Planck+WP+highL data, but does not lie in the $95\%$ CL of the
Planck+WP+highL+BICEP2. So, these potentials cannot provide the best fits with the current observations (see Fig.1 and Fig.2).
  \item For $V(\phi)=\sigma\phi$ and the given values
of $N$, the non-minimally kinetic coupled DGP model is well inside the $95\%$ CL Planck+WP+highL+BICEP2 data.
But, the evolution of tensor to scalar ratio  versus
scalar spectral index cannot provide the best fits with the current observations
(see Fig.3).

  \item For $V(\phi)=\sigma\phi^2$ and the given values
of $N$, the non-minimally kinetic coupled DGP model is well inside the $95\%$ CL Planck+WP+highL+BICEP2 data and
 the evolution of tensor to scalar ratio versus scalar spectral index  provides the best fits with the current observations
(see Fig.4).
 \item For $V(\phi)=\sigma\phi^3$ and the given values
of $N$, the non-minimally kinetic coupled DGP model is well inside the $95\%$ CL Planck+WP+highL+BICEP2 data and
 the evolution of tensor to scalar ratio versus
scalar spectral index  provides the best fits with the current observations
(see
 Fig.5).
\item For $V(\phi)=\sigma\phi^4$ and $N\leq60$, the non-minimally kinetic coupled DGP model is well inside the $95\%$ CL Planck+WP+highL+BICEP2 data. Since the number of e-folding
should be usually lager than 60 and because this potential cannot satisfy the slow-roll condition (i.e. $9\zeta H^2\dot{\phi}^{2}\ll V(\phi)$),  it is not a good potential for inflation in this model (see  Table II).

\item For $V(\phi)=\sigma\phi^5$ and the potentials with powers more than
5, one can show
that for the given values of $N$ the non-minimally kinetic
coupled DGP model is well inside the $95\%$ CL of the Planck+WP+highL data, but does not lie in the $95\%$ CL of the
Planck+WP+highL+BICEP2. Moreover, these potentials cannot satisfy the slow-roll condition (i.e $9\zeta H^2\dot{\phi}^{2}\ll V(\phi)$), hence   cannot be considered
as good potentials for inflation in this model (see Fig7 and Table
II).
\item For $V(\phi)=\sigma \ln(\phi)$ and $V(\phi)=\sigma e^{\pm\phi}$ and
the  given values of $N$, we  get the imaginary value of $\phi$ at the end of inflation (i.e. $\phi_e$). So, these potentials cannot be considered
as good potentials for inflation in this model (see Table II).

\item  For given scalar field potentials  with $n=1/2, 2/3, 1,2,3,4,5, 6$, $N=60$ and $\zeta=10^{-3}$ the non-minimally kinetic coupled DGP model lies inside the $95\%$ CL Planck+WP+highL data, and lies inside the $95\%$ CL Planck+WP+highL+BICEP2 data, for $n=1, 2$ (see Fig.8 and table I).

\item For given scalar field potentials with $n=1/2, 2/3, 1,2,3,4,5, 6$,  
$N=60$ and $\zeta=10^{-8}$, the non-minimally kinetic
coupled DGP model lies inside the $95\%$ CL Planck+WP+highL data for  $n=1/2, 2/3, 1,2$, and lies inside the $95\%$ CL Planck+WP+highL+BICEP2 data, for  $n=1,2,3,4,5, 6$  (see Fig.9 and table III). 

\item  For given scalar field potentials with $n=1/2, 2/3, 1,2,3,4,5, 6$, $N=60$ and $\zeta=10^{-10}$, the non-minimally kinetic
coupled DGP model lies inside the $95\%$ CL Planck+WP+highL data for  $n=1/2, 2/3, 1,2$, and lies inside the $95\%$ CL Planck+WP+highL+BICEP2 data, for  $n=1,2,3,4,5, 6$ (see Fig.10 and table IV). 
\end{itemize}

In conclusion, in the study of inflation using the non-minimally kinetic
coupled DGP model, we found that among the suggested potentials and coupling constants,    subject to the e-folding $N=60$ required by inflationary scenario,
the potentials $V(\phi)=\sigma\phi$,  $V(\phi)=\sigma\phi^2$ and $V(\phi)=\sigma\phi^3$  provide the best fits with both Planck+WP+highL data and Planck+WP+highL+BICEP2 data.
\acknowledgments
We would like to thank the anonymous referee whose useful comments much improved the presentation of this manuscript. This work has been supported financially by Research Institute
for Astronomy and Astrophysics of Maragha (RIAAM) under research project
NO.1/3720-6.


\end{document}